\documentclass[10pt,aps,prx,twocolumn,notitlepage,showpacs,superscriptaddress, %nofootinbib
]{revtex4-1}
\usepackage{amsthm,amsmath,amsopn,amsfonts,amssymb,verbatim,color}
\usepackage{graphicx}
\usepackage{subfigure}
\usepackage{bm}
\usepackage{epsfig,slashed}
\usepackage[T1]{fontenc}
\usepackage[colorlinks=true,citecolor=blue,linkcolor=blue,urlcolor=blue]{hyperref}
\usepackage{psfrag}
\DeclareMathOperator{\Res}{Res}
\DeclareMathOperator\Arctanh{Arctanh}
\DeclareMathOperator\arctanh{arctanh}

\begin{document}

\newcommand{\tr}{\mathop{\mathrm{Tr}}}
\newcommand{\bsigma}{\boldsymbol{\sigma}}
\newcommand{\re}{\mathop{\mathrm{Re}}}
\newcommand{\im}{\mathop{\mathrm{Im}}}
\renewcommand{\b}[1]{{\boldsymbol{#1}}}
\newcommand{\diag}{\mathrm{diag}}
\newcommand{\sign}{\mathrm{sign}}
\newcommand{\sgn}{\mathop{\mathrm{sgn}}}
\renewcommand{\c}[1]{\mathcal{#1}}
\renewcommand{\d}{\text{\dj}}
\newcommand{\red}{\textcolor{red}}

\newcommand{\mb}{\bm}
\newcommand{\ua}{\uparrow}
\newcommand{\da}{\downarrow}
\newcommand{\ra}{\rightarrow}
\newcommand{\la}{\leftarrow}
\newcommand{\mc}{\mathcal}
\newcommand{\bs}{\boldsymbol}
\newcommand{\lra}{\leftrightarrow}
\newcommand{\nn}{\nonumber}
\newcommand{\half}{{\textstyle{\frac{1}{2}}}}
\newcommand{\mf}{\mathfrak}
\newcommand{\MF}{\text{MF}}
\newcommand{\IR}{\text{IR}}
\newcommand{\UV}{\text{UV}}
\newcommand{\be}{\begin{equation}}
\newcommand{\ee}{\end{equation}}

\DeclareGraphicsExtensions{.png}

\title{Unconventional transport in low-density two-dimensional Rashba systems}

\author{Joel Hutchinson}
\email[electronic address: ]{jhutchin@ualberta.ca}
\affiliation{Department of Physics, University of Alberta, Edmonton, Alberta T6G 2E1, Canada}

\author{Joseph Maciejko}
\email[electronic address: ]{maciejko@ualberta.ca}
\affiliation{Department of Physics, University of Alberta, Edmonton, Alberta T6G 2E1, Canada}
\affiliation{Theoretical Physics Institute, University of Alberta, Edmonton, Alberta T6G 2E1, Canada}
\affiliation{Canadian Institute for Advanced Research, Toronto, Ontario M5G 1Z8, Canada}

\date\today

\begin{abstract}
Rashba spin-orbit coupling appears in 2D systems lacking inversion symmetry, and causes the spin-splitting of otherwise degenerate energy bands into an upper and lower helicity band. In this paper, we explore how impurity scattering affects transport in the ultra-low density regime where electrons are confined to the lower helicity band. A previous study has investigated the conductivity in this regime using a treatment in the first Born approximation. In this work, we use the full $T$-matrix to uncover new features of the conductivity. We first compute the conductivity within a semiclassical Boltzmann framework and show that it exhibits an unconventional density dependence due to the unusual features of the group velocity in the single particle dispersion, as well as quantized plateaus as a function of the logarithm of the electron density. We support this with a calculation using the Kubo formula and find that these plateaus persist in the full quantum theory. We suggest that this quantization may be seen in a pump-probe experiment. 
\end{abstract}

\pacs{
71.10.Ca,	% 	Electron gas, Fermi gas
71.70.Ej,	% 	Spin-orbit coupling, Zeeman and Stark splitting, Jahn-Teller effect
72.10.-d	% 	Theory of electronic transport; scattering mechanisms
}

\maketitle
%%%%%%%%%%%%%%%%%%%%%%%%%%%%%%
%%%%%%%%%%%%%%%%%%%%%%%%%%%%%%
\section{Introduction}
The generic consequence of broken inversion symmetry in quadratically dispersing bands with rotation and time-reversal symmetries is the development of Rashba spin-orbit coupling (SOC)~\cite{rashba1959, bychkov1984}. For a long time, the study of Rashba SOC was restricted to non-centrosymmetric crystals and two-dimensional (2D) quantum wells in heterostructures. It has since grown to ubiquity through advances in surface state measurement and manipulation~\cite{ast2007, gierz2009, Mirhosseini2010, Yaji2010}, synthetic SOC in ultra-cold atoms~\cite{huang2016}, and the recognition that local asymmetry can produce Rashba SOC even in centrosymmetric crystals~\cite{zhang2014}. 

In any case, the effect of Rashba SOC is to cause the dispersion to spin-split into two helicity bands as shown in Fig. \ref{fig:spectrum}. 
This splitting is bound to have profound effects on electron transport. Indeed, much of the spintronics industry, including the famous Datta-Das spin transistor~\cite{datta1990}, relies on the spin coherence of an electric current, and much theoretical effort has been focused on understanding whether scattering causes significant inter-band transitions~\cite{walls2006, manchon2015, khaetskii2006}. Only recently has attention been paid to the low-density regime, where elastic scattering causes intra-band transitions within an annular Fermi sea~\cite{jo2017}. Many-body phases in this regime are particularly interesting due to the highly degenerate ring of momentum states at the band bottom~\cite{takei2012, wang2010, berg2012, ruhman2014, silvestrov2014}, and its corresponding singular density of states~\cite{cappelluti2007}. In this paper we focus on the effects of this low-density ring on impurity scattering and thereby transport in 2D Rashba materials. It has been recognized that the DC conductivity is a nonlinear function of the density in this regime~\cite{brosco2016, brosco2017, brosco2017_2}. This paper extends the work of these references to include non-perturbative scattering effects that arise at ultra-low densities. The dynamics of single-electron scattering in this regime was explored in Ref.~\cite{hutchinson2016} by examining the $S$-matrix for specific impurity potentials. In Ref.~\cite{hutchinson2017} it was shown that the corresponding $T$-matrix takes on a universal and unconventional form for any circularly symmetric, spin-independent potential. This form leads to plateaus in the scattering cross-section as a function of the logarithm of the energy.

In this paper we demonstrate that Rashba transport at low densities exhibits unusual features including quantized conductivity. The outline of the paper is as follows. After some brief preliminaries to establish notation with regards to the free electron spectrum,  we begin with Sec.~\ref{sec:Boltz}, a Boltzmann calculation of the conductivity within linear response. The key difference between this and a standard Boltzmann treatment is the use of the full $T$-matrix derived in Ref.~\cite{hutchinson2017}. This is qualitatively different from previous treatments done in the first Born approximation where the conductivity is found to smoothly decay to zero as the density is decreased~\cite{brosco2016, brosco2017, brosco2017_2}. We first focus on the zero temperature DC and AC conductivities (Sec.~\ref{subsec:zeroT}), before generalizing to finite temperature (Sec.~\ref{subsec:finiteT})  where we allow the chemical potential to vary through and below the conduction band. Such a treatment describes Rashba semiconductors and allows us to outline a possible experimental realization of the unique conductivity features we uncover in this section. Sec.~\ref{sec:SCFBA} goes beyond the Boltzmann approach to include quantum corrections within a self-consistent full Born approximation. Again, the difference between this and previous work is that the self-energy is computed self-consistently from the full $T$-matrix. We first focus on the single-particle Green's function, self-energy and density of states (Sec.~\ref{subsec:single}) before employing this approach in a calculation of the conductivity using the Kubo formula (Sec.~\ref{subsec:Kubo}).
%%%%%%%%%%%%%%%%%%%
\subsection{Preliminaries}\label{subsec:prelim}

\begin{figure}[t]
	\centering
	\includegraphics[width=\columnwidth]{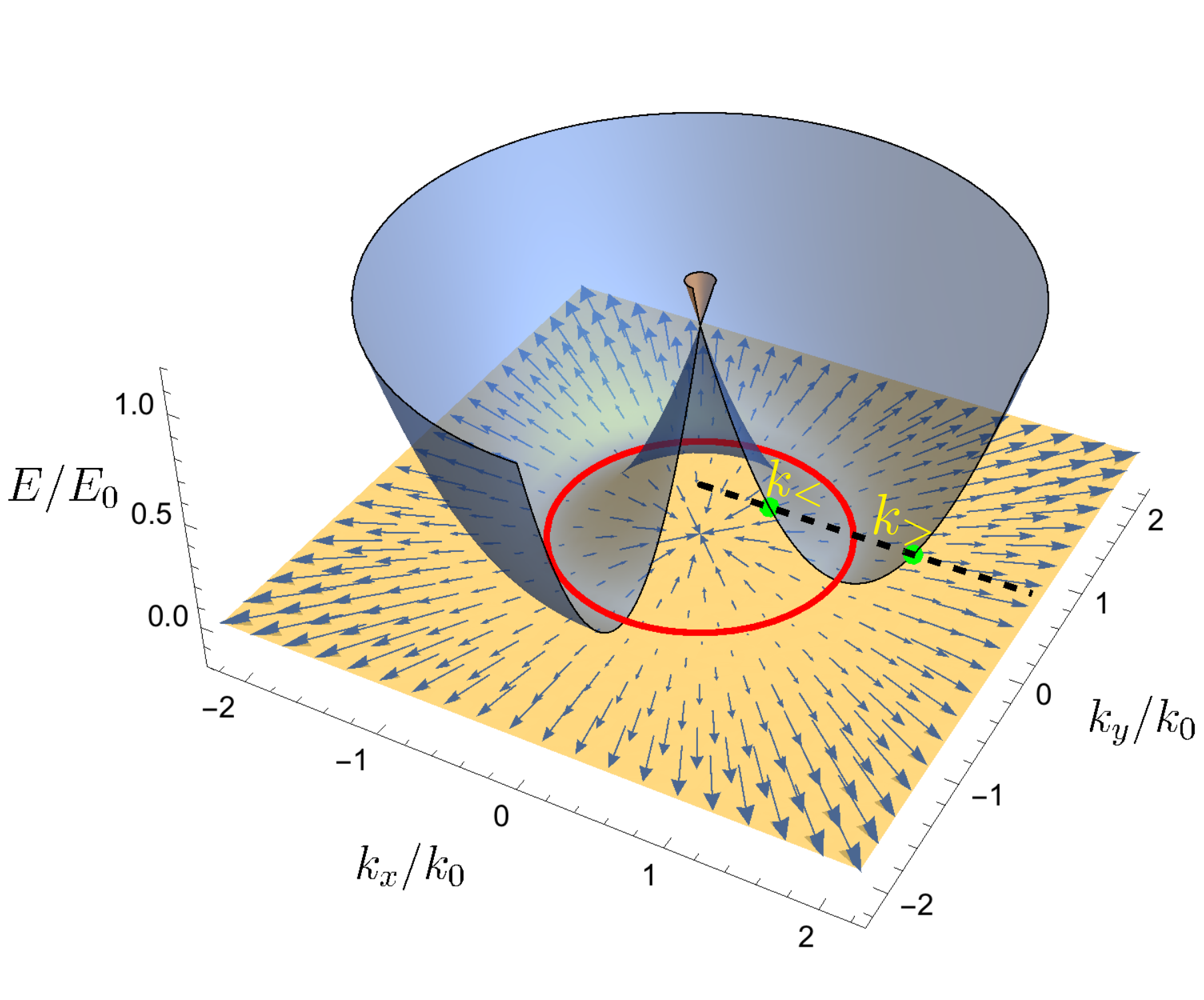}
\caption{Lower helicity dispersion $\xi_k^-+\mu$ (in blue). The plane beneath it shows the corresponding group velocity vector field which changes from pointing inward to pointing outward at the band minimum (shown in red). At a generic energy below the Dirac point (dashed line), there are two available rings of states with wavenumber $k_<$ and $k_>$.}\label{fig:spectrum}
\end{figure}

The single-particle 2D Rashba Hamiltonian in the continuum limit is given by,
\be\label{eq:HRashba}
H(\b{k})=\frac{\b{k}^2}{2m}+\lambda\hat{\b{z}}\cdot(\b{\sigma}\times\b{k}),
\ee
where $\b{k}=(k_x,k_y,k_z)$ is the momentum, $\b{\sigma}=(\sigma^x, \sigma^y, \sigma^z)$ is a vector of Pauli matrices, $m$ is the electron mass and $\lambda$ is the Rashba coupling. This Hamiltonian admits a spectrum with two helicity bands that meet at a Dirac point as shown in Fig.~\ref{fig:spectrum}. This spectrum, measured with respect to the chemical potential $\mu$ is
\be
\xi_k^\pm=\frac{k^2}{2m}\pm\lambda k-\mu+E_0.
\ee
Throughout this paper we focus exclusively on Fermi energies $E_F$ below the Dirac point, where only one helicity band is present. The energy difference between the Dirac point and the band bottom is $E_0\equiv\frac{1}{2}m\lambda^2$. The band bottom is a ring at finite momentum $k_0\equiv m\lambda$. We work with a shifted spectrum where $E_f=0$ occurs at the band bottom. It is convenient to use the dimensionless energy parameter $\delta\equiv\sqrt{E/E_0}$.
At any Fermi energy %$-E_0<E_f<0$ 
$0<E_f<E_0$, the Fermi sea is defined by an annulus with two wavenumbers designated $k_{\gtrless}\equiv k_0(1\pm\delta)$. The two states differ in the value $s_\mu\equiv\text{sgn}(k_\mu-k_0)=\pm1$. Equivalently, we can note that their group velocities are oppositely oriented: $s_\mu=\text{sgn}(\b{v}_\mu\cdot\b{k}_\mu)$.

%fig. 1
%%%%%%%%%%%%%%%%%%%%%%%%%%%%%%
%%%%%%%%%%%%%%%%%%%%%%%%%%%%%%%
\section{Semiclassical Boltzmann Transport}\label{sec:Boltz}
\subsection{Zero temperature}\label{subsec:zeroT}
\subsubsection{DC Conductivity}

In the Boltzmann approach, we look for the deviation, $n_{\mu}(\phi_{\b{k}},E)$, from the equilibrium distribution, that satisfies the translation invariant Boltzmann equation. In the linear response regime, with an impurity density $n_i$ and an electric field $\b{\mathcal{E}}$ oscillating at frequency $\omega$, this equation reads
\begin{eqnarray}
i\omega n_{\b{k}_\mu}&=&-e\b{\mathcal{E}}\cdot\b{\nabla}_{\b{k}_\mu}n^0_{\b{k}_\mu}\nonumber\\
&&+n_{i}\sum_\nu\int_0^{2\pi}d\phi_{\b{k}}'\;W^{\phi_{\b{k}}\phi_{\b{k}}'}_{\mu\nu}[n_\mu(\phi_{\b{k}},E)-n_\nu(\phi_{\b{k}}',E)].\nonumber\\
\label{eq:boltz}
\end{eqnarray}
We will set $\omega=0$ for now. Here, $\mu$ and $\nu$ indicate the $>$,$<$ states described in Sec.~\ref{subsec:prelim}, and $\phi_{\b{k}}$ is the in-plane angle of the corresponding wavevector $\b{k}_\mu$~\footnote{The single-particle state is specified either by $\b{k}$, or by $\{E,\mu,\phi_{\b{k}}\}$, but we will often use the redundant notation $\b{k}_\mu$ for clarity.}.
The equilibrium distribution $n^0_{\b{k}_\mu}$ is given by the Fermi function $f(E-\mu)$:
\be
\b{\nabla}_{\b{k}_\mu}n^0_{\b{k}_\mu}=\frac{\partial f}{\partial E}\b{\nabla}_{\b{k}_\mu}\xi_{k}=\frac{k_0\delta}{m}\frac{\partial f}{\partial E}s_\mu\hat{k},
\ee  
where $\hat{k}=\b{k}/|\b{k}|$. The matrix $n_iW_{\mu\nu}^{\phi_{\b{k}}\phi_{\b{k}}'}$ is the elastic scattering rate between state $|E,\nu,\phi_{\b{k}}'\rangle$ and $|E,\mu,\phi_{\b{k}}\rangle$ determined from Fermi's golden rule. For circularly symmetric impurity potentials, it depends only on the difference $\tilde{\phi}_k\equiv\phi_{\b{k}}-\phi_{\b{k}}'$, and the corresponding $T$-matrix $T^{\b{k}_\mu\b{k}_\nu}$ admits an expansion in circular harmonics, which at low energies is independent of the magnitude of the wavevectors~\cite{hutchinson2017}:
\begin{eqnarray}
W_{\mu\nu}^{\phi_{\b{k}}\phi_{\b{k}}'}&=&|T^{\b{k}_\mu\b{k}_\nu}|^2g_\nu(E)\\
&=&\frac{m}{2\pi\delta}\bigg |\sum_{l=-\infty}^\infty T^l(E)e^{il\tilde{\phi}_k}\bigg |^2(1+s_\nu\delta),\label{eq:fermirule}
\end{eqnarray}
where we have used the density of states in the $\nu$ channel,
\be\label{eq:dos}
g_\nu(E)=\int \frac{d^2\b{k}}{(2\pi)^2}\delta(E-\xi_{k_\nu})=\frac{m}{2\pi\delta}(1+s_\nu\delta).
\ee
In Appendix \ref{app:symm}, we show that the scattering rate above satisfies detailed balance. Using our non-perturbative solution for the low-energy $T$-matrix~\cite{hutchinson2017} will allow us to go well beyond the usual perturbative treatments in the Born approximation.

Next, we choose the following ansatz for the distribution function,
\be
n_{\mu}(\phi_{\b{k}},E)=\sum_\nu\Gamma^{-1}_{\mu\nu}e\b{\mathcal{E}}\cdot\b{\nabla}_{\b{k}_\nu}n^0_{\b{k}_\nu},
\ee
where $\Gamma$ is a $2\times2$ matrix to be determined.
Substituting this into equation \eqref{eq:boltz} and integrating over $\phi_{\b{k}}$ gives a matrix equation in the $>$, $<$ basis,
\begin{eqnarray}
&&\bigg[\bigg (\frac{1}{\tau}\Gamma^{-1}-\mathbb{I}\bigg )-\frac{(1/\tau-1/\tau^{\text{tr}})}{2}\begin{pmatrix}1+\delta & 1-\delta\\ 1+\delta & 1-\delta\end{pmatrix}\Gamma^{-1}\bigg]\begin{pmatrix} 1 \\ -1 \end{pmatrix}\nonumber\\
&&=0.\label{eq:boltz2}
\end{eqnarray}
In analogy with the conventional spin-degenerate system, we define the energy-dependent lifetime $\tau$ and transport time $\tau^{\text{tr}}$ as
\begin{eqnarray}
\frac{1}{\tau}&\equiv&\frac{n_im}{\pi\delta}\int_0^{2\pi}d\tilde{\phi}_k\bigg|\sum_{l=-\infty}^\infty T^l(E)e^{il\tilde{\phi}_k}\bigg|^2,\label{eq:tau}\\
\frac{1}{\tau^\text{tr}}&\equiv&\frac{n_im}{\pi\delta}\int_0^{2\pi}d\tilde{\phi}_k(1-\cos\tilde{\phi}_k)\bigg|\sum_{l=-\infty}^\infty T^l(E)e^{il\tilde{\phi}_k}\bigg|^2.\label{eq:tautr}\nonumber\\
\end{eqnarray}
In Sec.~\ref{subsec:single} we will see that these definitions are consistent with the lifetime derived from the self-energy.

Equation \eqref{eq:boltz2} is readily solved by the matrix
\be
\Gamma^{-1}=
\begin{pmatrix} \tau^\text{tr} & (1-\delta)(\tau^\text{tr}-\tau)\\
(1+\delta)(\tau^\text{tr}-\tau) & \tau^\text{tr}
\end{pmatrix},
\ee
from which we get the distribution function,
\be\label{eq:ansatz1}
n_\mu(\phi_{\b{k}},E)=\frac{k_0\delta}{m}\frac{\partial f}{\partial E}e\b{\mathcal{E}}\cdot\hat{k}s_\mu(\tau+s_\mu\delta(\tau^\text{tr}-\tau)).
\ee
From this, we calculate the current,
\begin{eqnarray}
\b{J}&=&-e\sum_\mu\int dE\int \frac{d\phi_{\b{k}}}{2\pi}g_\mu(E)n_\mu(\phi_{\b{k}},E)\b{\nabla}_{\b{k}_\mu}\xi_k^-\\
&=&-\frac{e^2k_0^2}{2\pi m}\int dE\delta\frac{\partial f}{\partial E}(\tau +\delta^2(\tau^\text{tr}-\tau))\b{\mathcal{E}}.\label{eq:current}
\end{eqnarray}
Taking the zero-temperature limit, we get the DC conductivity
\be
\sigma_\text{DC}=\frac{e^2}{2\pi}\frac{k_0^2\delta}{m}(\tau +\delta^2(\tau^\text{tr}-\tau)),
\ee
where it is understood that the energies in this expression are evaluated at the Fermi level.
We may write this in terms of the electron density using
\be
n=\frac{k_0^2}{\pi}\delta,
\ee
which follows from \eqref{eq:dos}. Thus,
\be\label{eq:dccond}
\sigma_\text{DC}=\frac{e^2}{2}\frac{n}{m}\bigg[\tau +\bigg(\frac{n}{n_0}\bigg)^2(\tau^\text{tr}-\tau)\bigg],
\ee
where $n_0\equiv k_0^2/\pi$ is the density at the Dirac point.

There are several important features to note about \eqref{eq:dccond}. First, recall that $\tau$ and $\tau^\text{tr}$ both depend on the density  through $\delta$ and $T^l(E_F)$ in \eqref{eq:tau} and \eqref{eq:tautr}, so the conductivity is a highly non-linear function of the density. Second, it reproduces the Drude conductivity
\be
\sigma^{\rm Drude}=\frac{e^2}{2}\frac{n}{m}\tau^\text{tr},
\ee
only at the Dirac point where $n\rightarrow n_0$, though our low-energy expression for the transport time \eqref{eq:tautr} is not accurate in this regime. What is special for transport about the Dirac point is that there is only one channel ($k_>$) with a non-vanishing density of states, and a group velocity parallel to $\b{k}$, just as in a typical parabolic dispersion for a single fermion species (giving the $1/2$ in the Drude conductivity). In the opposite limit, $n\rightarrow0$, we see that it is the \emph{lifetime}, not the transport time that governs the conductivity
\be\label{eq:BoltzmannCondLowDensity}
\sigma_\text{DC}(n\rightarrow0)=\frac{e^2}{2}\frac{n}{m}\tau.
\ee   
This is because in this limit, one has $k_>\approx k_<$, so that these two channels have approximately the same phase space for scattering. But since they have oppositely directed group velocities, scattering through an angle $\tilde{\phi}_k=\pi$ is just as likely to result in forward scattering as scattering through an angle $\tilde{\phi}_k=0$~\cite{hutchinson2016, hutchinson2017erratum}.

We will see shortly that the unusual density dependence of the \emph{full} $T$-matrix results in novel features in the conductivity. However, we first consider the first Born approximation as was done in Ref.~\cite{brosco2016}, for which the $T$-matrix is given by a spin-independent constant potential transformed to the helicity basis: 
\begin{eqnarray}
\bigg|\sum_{l=-\infty}^\infty T^l(E)e^{il\tilde{\phi}_k}\bigg|^2&=&|v_0\eta^\dagger_-(\phi_{\b{k}})\eta_-(\phi_{k'})|^2\\
&=&\frac{v_0^2}{2}(1+\cos\tilde{\phi}_k),
\end{eqnarray}
where $\eta_-(\phi_{\b{k}})=\frac{1}{\sqrt{2}}\begin{pmatrix} 1 \\ ie^{i\phi_{\b{k}}}\end{pmatrix}$ is the negative-helicity eigenspinor of the Hamiltonian \eqref{eq:HRashba}.
In this case, the lifetime and transport time become
\be\label{eq:tau1ba}
\frac{1}{\tau}=n_imv_0^2\frac{n_0}{n}=\frac{2}{\tau^\text{tr}},
\ee
and the conductivity is
\be
\sigma_\text{DC}^\text{1BA}=\frac{e^2}{2n_im^2v_0^2}\frac{n^2}{n_0}\bigg[1+\bigg(\frac{n}{n_0}\bigg)^2\bigg],
\ee
in agreement with~\cite{brosco2016}.

\begin{figure}[t]
	\centering
	\includegraphics[width=\columnwidth]{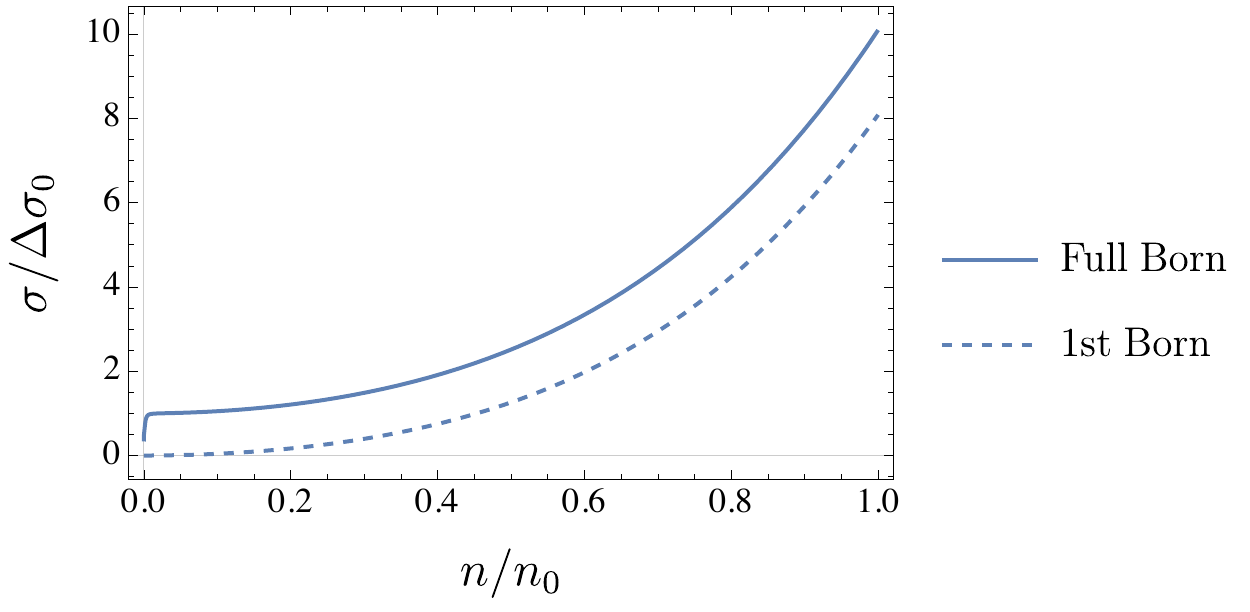}
	\includegraphics[width=\columnwidth]{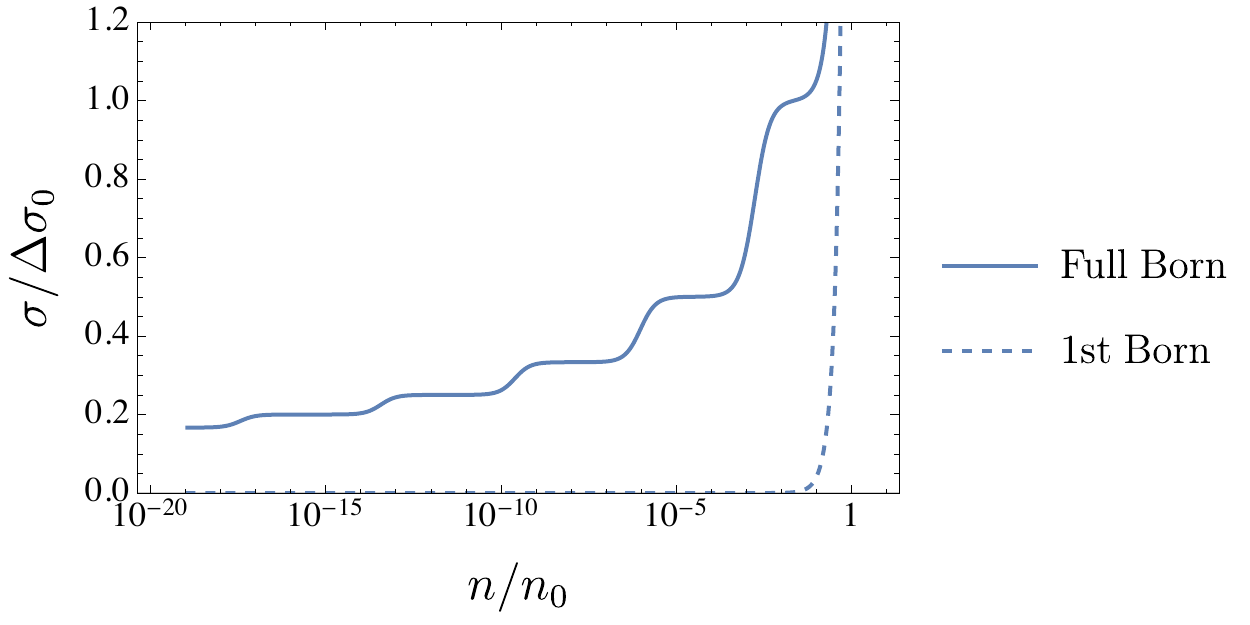}
\caption{DC conductivity divided by $\Delta\sigma_0\equiv\frac{e^2}{8\hbar}\frac{n_0}{n_i}$ vs electron density on a linear (top) and log (bottom) scale. The solid lines indicate the Boltzmann result using the full Born $T$-matrix. The dashed lines show the corresponding first Born approximation for the conductivity. A delta-shell impurity potential is used with the parameters $mv_0R^2=1$ and $R=0.1/k_0$.}\label{fig:DCcond}
\end{figure}

In Fig.~\ref{fig:DCcond}, we plot the density dependence of the DC conductivity for the case where the impurity potentials are modeled by $\delta$-function shells $V(r)=v_0R\delta(r-R)$. The single-impurity Rashba $T$-matrix was computed non-perturbatively in Ref.~\cite{hutchinson2017} in the low-energy limit, for any circularly symmetric impurity potential. It has the form
\be
T^l(E)\approx\frac{1}{m}\frac{\delta^*_l}{1+i\delta_l^*/\delta},
\ee
where $\delta_l^*$ parameterizes the $l$th circular harmonic of the matrix element of the impurity potential between two states at the band bottom: 
\be
\delta^*_l\equiv\frac{m}{2}(V^l(k_0,k_0)+V^{l+1}(k_0,k_0)).
\ee
The specific choice of impurity potential makes no qualitative difference, as long as rotational symmetry is maintained. The $\delta$-shell potential contains two independent parameters, the impurity strength $v_0$ and radius $R$. Varying these parameters simply changes the scales in Fig.~\ref{fig:DCcond}. Increasing the impurity strength decreases the conductivity everywhere, and increasing its radius shifts the plateaus of the bottom panel to higher densities. Indeed, one can find a quantitative estimate of the effect of these parameters for a given potential. In the long-wavelength limit $k_0R\ll1$, we have
\begin{eqnarray}
\delta^*_l\approx\frac{mv_0R^2}{2(|l|!)^2}\bigg(\frac{(k_0R)^2}{4}\bigg)^{|l|}\;\;&& \text{($\delta$-shell)}\label{eq:dlstarDeltaShell}\\
\delta^*_l\approx\frac{mv_0R^2}{4|l|!|l+1|!}\bigg(\frac{(k_0R)^2}{4}\bigg)^{|l|}\;\;&& \text{(hard disk)}. 
\end{eqnarray}
With this in mind, we will maintain the same parameter values throughout the paper.
%fig. 2

Figure \ref{fig:DCcond} constitutes the main result of this paper, so we pause to flesh out the salient observations contained within. First, as seen in the top panel, the conductivity does not decay smoothly to zero with decreasing density, unlike the prediction from the first Born approximation. In fact, measurements at low densities might lead one to believe there is a finite residual conductivity as $n\rightarrow0$. This is not physical, and indeed is not the case as shown in the lower panel, where we see that the conductivity goes through a series of steps (on a logarithmic scale) to reach zero at $n=0$. On such a scale, the conductivity is quantized, with plateaus given by the values
\be\label{eq:plateaus}
\sigma_\text{DC}=\bigg(\frac{n_0}{n_i}\bigg)\frac{e^2}{8\hbar l},\;\;l=1,2,3,\dots,
\ee
and transitions between plateaus occurring at $n/n_0=\delta_l^*$.
Although the conductivity plateaus depend on material parameters such as the impurity concentration and the spin-orbit coupling strength, the ratio of any two conductivity plateaus is a pure rational number, independent of all such parameters. The origin of these plateaus is the energy scale separation of the different circular harmonic contributions to the $T$-matrix, discussed in detail in~\cite{hutchinson2016, hutchinson2017}. Briefly, let us give a physical explanation of this origin. We consider here finite impurity potentials which require an infinite number of angular components to describe them (as opposed to a delta-function impurity which only has an $s$-wave component). Each of these components are well separated in magnitude because they are controlled by Bessel functions of different orders (i.e. $J_l(k_0R)$). Due to interference between the $k_<$ and $k_>$ scattering states, the scattered wavefunction becomes more and more quasi-one-dimensional as the energy is lowered, resembling a plane wave at the band bottom. This plane wave is composed of equal contributions of all angular components which are turned on at successively lower scattering energies, essentially when $\delta\approx\delta_l^*$. Each time this happens, a new angular channel contributes to the scattering and the conductivity drops by a quantized amount.

%In any scattering problem, each partial wave of the scattered wavefunction is imparted its own phase shift $\delta_l$ by the corresponding angular component of the scattering potential. We consider here finite impurity potentials which require an infinite number of angular components to describe them (as opposed to a delta-function impurity which only has an $s$-wave component). Each of these phase shifts are well separated in magnitude because the corresponding potential components are controlled by Bessel functions of different orders. In conventional scattering, this means that the $s$-wave phase shift is parametrically larger than higher partial waves. This is true in the Rashba system as well, but the situation is more complicated. In the Rashba system, there are two states of different wavelengths available for outgoing scattered waves ($k_>$ and $k_<$). These waves combine to produce a beat pattern with wavenumber $2k_0\delta$. This envelope wavefunction comes with its own phase shift that is independent of partial wave number. As the scattering energy is lowered, one by one the individual phase shifts of each partial wave collapse onto this universal energy-dependent value. Each time this happens, a drop in the conductivity is observed as one more partial-wave channel is contributing to scattering.
%%%%%%%%%%%%%%%%%%%%%%%%%
\subsubsection{AC Conductivity}
\begin{figure}[t]
	\centering
	\includegraphics[width=\columnwidth]{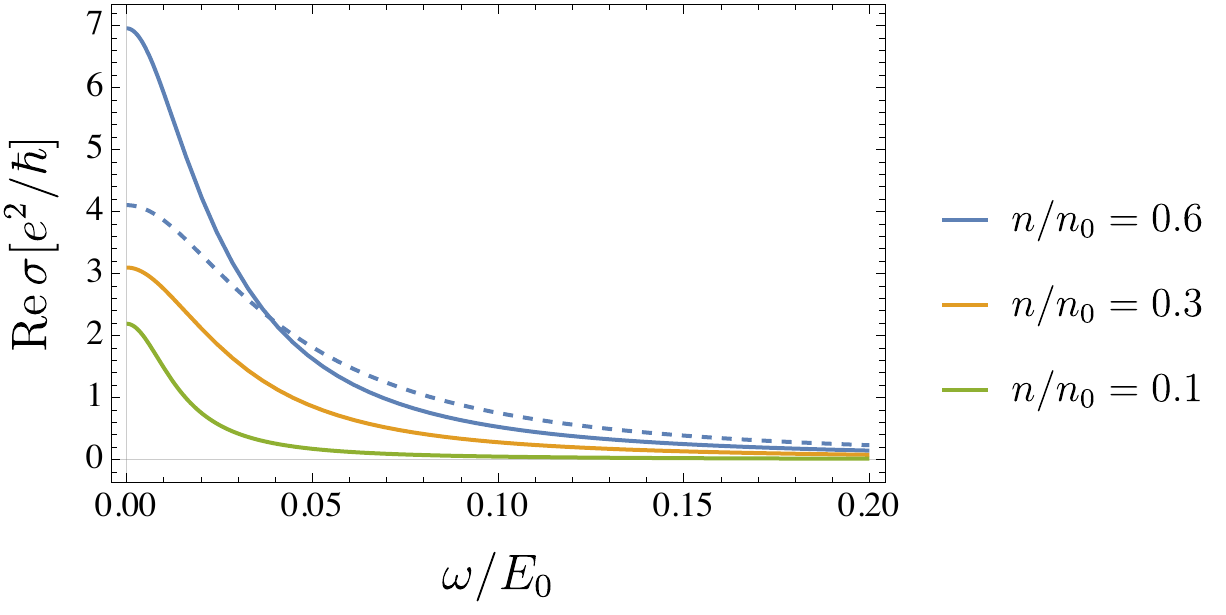}
	\includegraphics[width=\columnwidth]{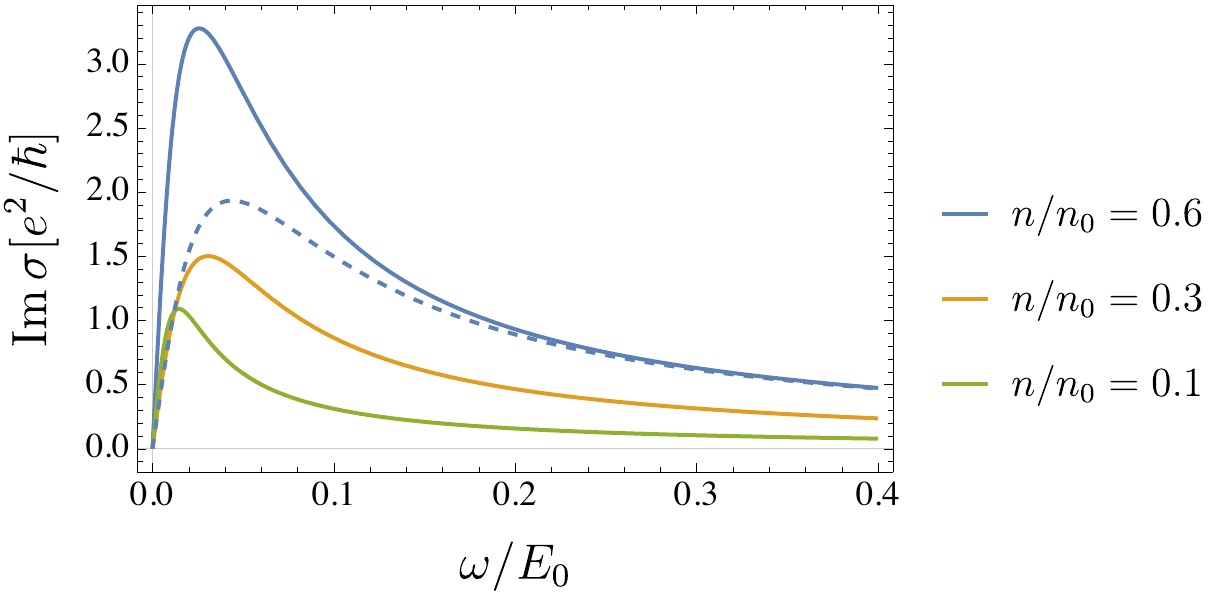}
\caption{Real (top) and imaginary (bottom) parts of the semiclassical Boltzmann AC conductivity vs frequency for various values of the electron density with impurity density set to $n_i=0.06n_0$. The dashed line shows the corresponding ($n=0.6n_0$) first Born approximation result for the conductivity, found by inserting \eqref{eq:tau1ba} into \eqref{eq:ACcond}. %A delta-shell impurity potential is used with the parameters $mv_0R=1$ and $R=0.1/k_0$.
}\label{fig:ACcond}
\end{figure}

Retaining the frequency dependence in \eqref{eq:boltz} allows us to compute the AC conductivity as well. 
%In this case, we modify our ansatz for the electron distribution \eqref{eq:ansatz1} to
%\be
%n_{\b{k}_\mu}(\omega)=-A(\omega)\sum_\nu(i\omega-\Gamma)^{-1}_{\mu\nu}e\b{\mathcal{E}}\cdot\b{\nabla}_{\b{k}_\nu}n^0_{\b{k}_\nu}.
%\ee
%In matrix form, the Boltzmann equation now reads
One can readily check that the AC Boltzmann equation is solved by
\be
n_\mu(\phi_{\b{k}},E)=\frac{k_0\delta}{m}\frac{\partial f}{\partial E}e\b{\mathcal{E}}\cdot\hat{k}s_\mu\bigg(\frac{\tau(1-i\omega\tau^\text{tr})+s_\mu\delta(\tau^\text{tr}-\tau)}{(1-i\omega\tau^\text{tr})(1-i\omega\tau)}\bigg),
\ee
from which we get the current
\be
\b{J}=-\frac{e^2k_0^2}{2\pi m}\int dE\delta\frac{\partial f}{\partial E}\bigg(\frac{\tau(1-i\omega\tau^\text{tr})+\delta^2(\tau^\text{tr}-\tau)}{(1-i\omega\tau^\text{tr})(1-i\omega\tau)}\bigg)\b{\mathcal{E}},
\ee
and the zero temperature conductivity
\be\label{eq:ACcond}
\sigma(\omega)=\frac{e^2}{2}\frac{n}{m}\bigg(\frac{\tau(1-i\omega\tau^\text{tr})+(n/n_0)^2(\tau^\text{tr}-\tau)}{(1-i\omega\tau^\text{tr})(1-i\omega\tau)}\bigg),
\ee
shown in Fig.~\ref{fig:ACcond}.

\begin{figure}[t]
	\centering
	\includegraphics[width=\columnwidth]{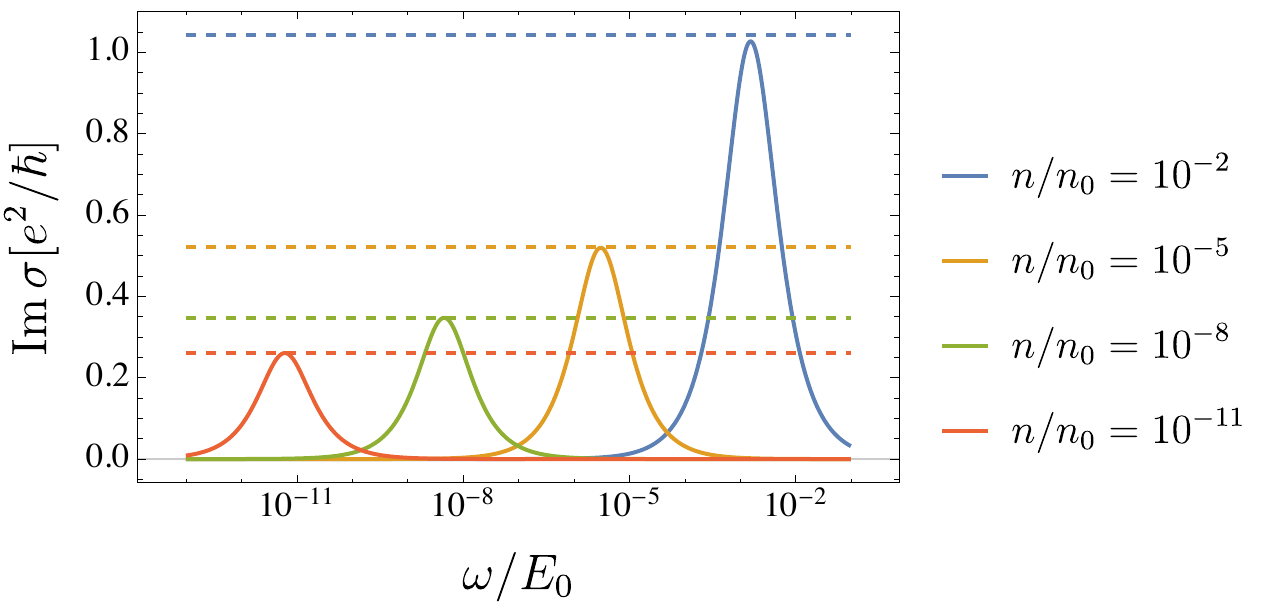}
\caption{Imaginary part of the semiclassical Boltzmann AC conductivity vs frequency for various values of the electron density on a log scale with impurity density set to $n_i=0.06n_0$. The dashed lines show the quantization of the peaks given by $n_0/n_i(\frac{1}{16l})$ for $l=1,2,3,4$.%  A delta-shell impurity potential is used with the parameters $mv_0R=1$ and $R=0.1/k_0$.
}\label{fig:ACcondQuant}
\end{figure}

As we saw in the DC case, the conductivity takes the Drude form in the two limits $n\rightarrow n_0$ and $n\rightarrow 0$, dependent on $\tau^{\rm tr}$ and $\tau$ respectively. As $n\rightarrow n_0$,
\be
\sigma(\omega)\rightarrow\frac{e^2}{2}\frac{n}{m}\frac{\tau^\text{tr}}{1-i\omega\tau^\text{tr}}=\sigma^\text{Drude}(\omega),
\ee
while in the opposite limit $n\rightarrow0$,
\be
\sigma(\omega)\rightarrow\frac{e^2}{2}\frac{n}{m}\frac{\tau}{1-i\omega\tau}.
\ee
Thus, as the density is lowered, the width of the Drude peak decreases from $1/\tau^\text{tr}$ to $1/\tau$. At low densities, the height of the peak in the imaginary part of the AC conductivity at $\omega=1/\tau$ becomes quantized since 
\be
\im\sigma(1/\tau)\approx\frac{e^2}{2}\frac{n}{m}\frac{\tau}{2}=\frac{\sigma_{\rm{DC}}}{2},
\ee
where $\sigma_{\rm{DC}}$ has the plateaus in \eqref{eq:plateaus}, shown in Fig.~\ref{fig:ACcondQuant}.
%fig. 3

%fig. 4
%%%%%%%%%%%%%%%%%%%%%%%%%%%%%%%%%%%
%%%%%%%%%%%%%%%%%%%%%%%%%%%%%%%%%%%%%%%%%
\subsection{Finite temperature}\label{subsec:finiteT}
We now consider the effect of finite temperature on transport in a Rashba semiconductor. Several materials exist with large Rashba splitting of order $E_0\sim0.1\rm{eV}$, including BiTeI and CH$_3$NH$_3$PbI$_3$~\cite{bahramy2017, ishizaka2011, kepenekian2017, zheng2015, zhao2016}. For example, BiTeI possesses 2D surface conduction and valence bands with a large Rashba splitting; the Fermi level can be adjusted between these two bands by changing the termination layer~\cite{crepaldi2012}. In the following, however, we ignore material-specific details and consider a simplified model of a semiconductor with a Fermi level close to the bottom of a 2D Rashba-split conduction band.

\begin{figure}[t]
	\centering
	\includegraphics[width=\columnwidth]{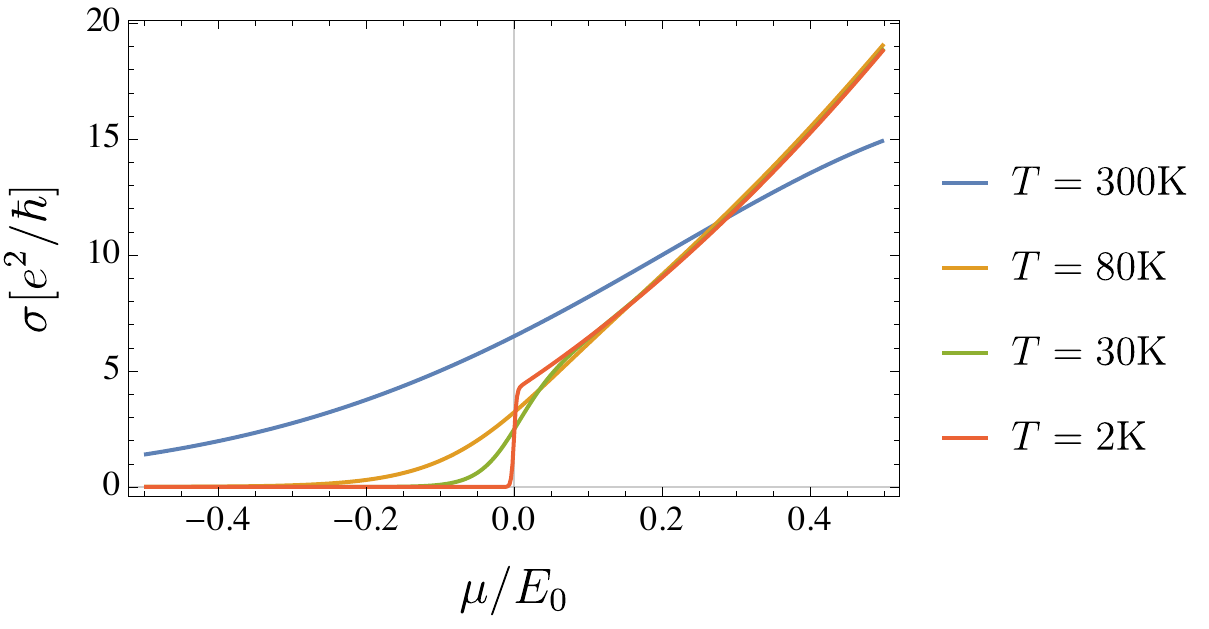}
	\includegraphics[width=\columnwidth]{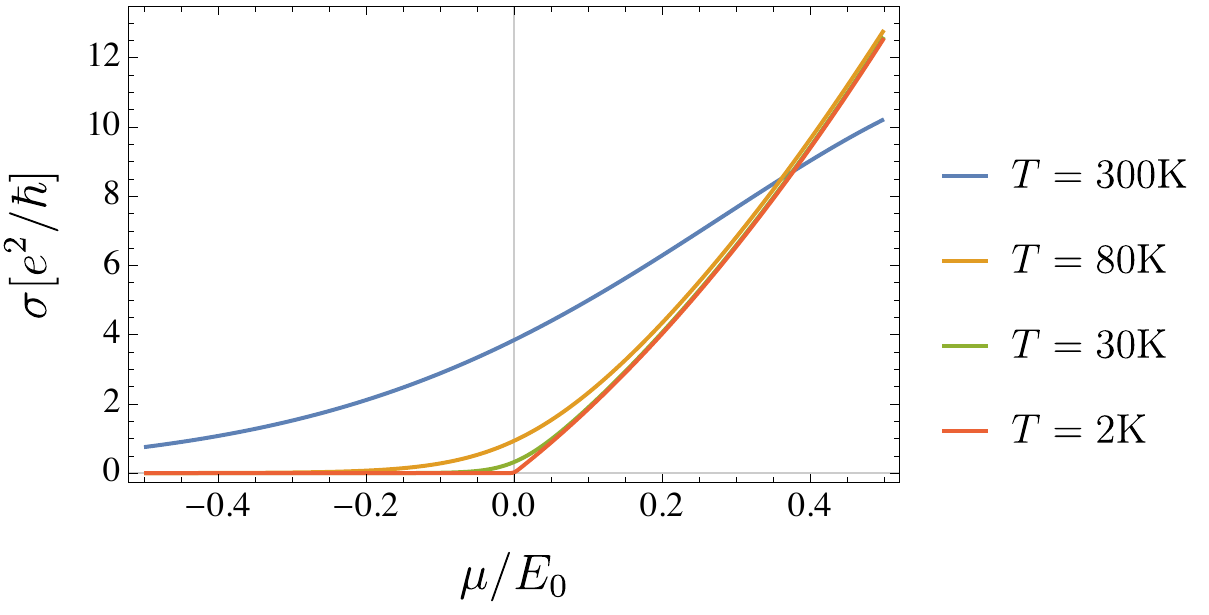}
\caption{Semiclassical Boltzmann DC conductivity vs chemical potential for various temperatures, with impurity density set to $n_i=0.03n_0$. The top panel shows the results computed from the full $T$-matrix, while the bottom panel shows the first Born approximation.
}\label{fig:chempot}
\end{figure}

Returning to \eqref{eq:current}, we retain the temperature dependence via 
\be
\frac{\partial f}{\partial E}=-\frac{\beta e^{\beta (E-\mu)}}{(e^{\beta (E-\mu)}+1)^2}.
\ee
Furthermore, we assume that $|E_0-\mu|\gg k_BT$, so that the upper helicity band does not contribute to the integrand. The DC conductivity, 
\be\label{eq:condfiniteT}
\sigma_\text{DC}\approx\frac{e^2k_0^2\beta}{2\pi m}\int_{0}^{E_0}dE\;\delta\frac{e^{(E-\mu)\beta}}{(e^{\beta(E-\mu)}+1)^2}(\tau +\delta^2(\tau^\text{tr}-\tau)),
\ee 
is then computed numerically as a function of the chemical potential. The result is shown in the top panel of Fig.~\ref{fig:chempot}, where we see that the sharp zero temperature drop that occurs at the band bottom ($\mu=0$) maintains some weight at finite temperatures. The magnitude of this drop is determined by the impurity density
\be\label{eq:DeltaSigma}
\Delta\sigma(\mu=0)=\Delta\sigma_0\equiv\frac{e^2}{8\hbar}\frac{n_0}{n_i}.
\ee
This is in contrast to the prediction from the first Born approximation shown in the bottom panel of Fig.~\ref{fig:chempot}, which produces a conductivity that smoothly goes to zero as the chemical potential is lowered.
%fig. 5

\begin{figure}[t]
	\centering
	\includegraphics[width=\columnwidth]{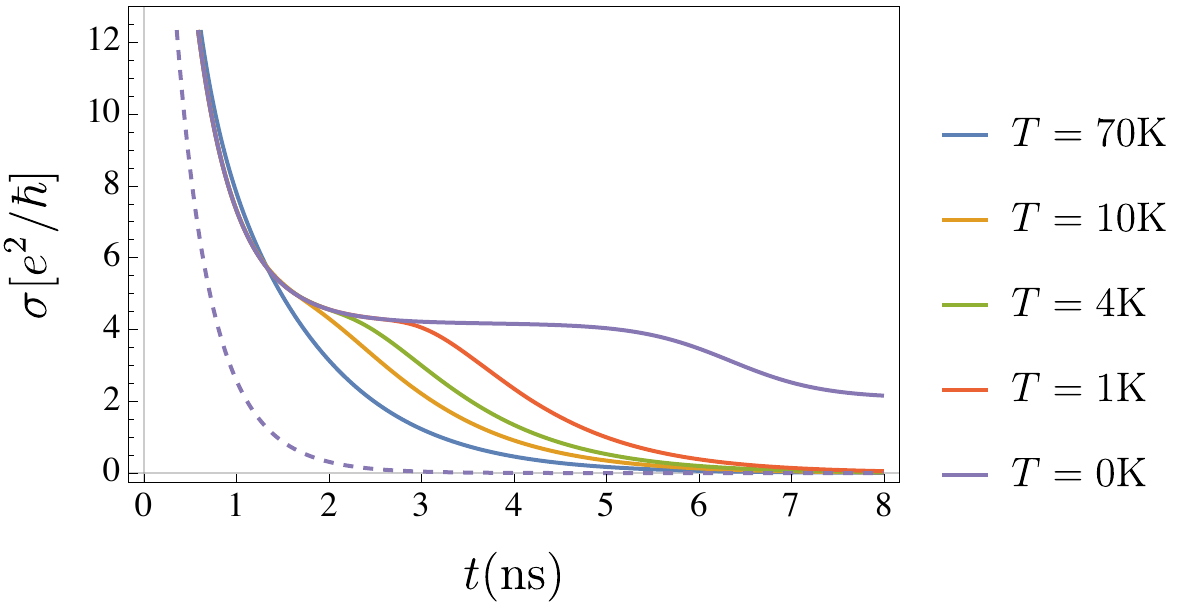}
\caption{Semiclassical Boltzmann DC conductivity vs delay time in a pump-probe measurement at various temperatures, with impurity density set to $n_i=0.03n_0$, and recombination time $\tau_n=1$ ns. The dashed line shows the ($T=0$) prediction from the first Born approximation.
}\label{fig:delay}
\end{figure}

The conductivity plateaus seen in the zero temperature case are hidden in the sharp drop near $\mu=0$. In a real material, doping offers crude control over the chemical potential, and one might be skeptical that the quantization seen as a function of the logarithmic changes in the density (or alternatively, the logarithmic changes in the chemical potential) could ever be observed. As one potential means of overcoming this we propose the use of pump-probe measurements. This technique has been used to study transport properties of systems with large Rashba splitting before~\cite{Valverde2015}. In such an experiment, a (typically THz) pump laser pulse is used to excite carriers from the valence to conduction band. These carriers quickly establish a quasi-equilibrium and a corresponding chemical potential $\mu>0$, on a time scale ($\sim10^{-15}$ s) much smaller than the typical recombination time $\tau_n\sim10^{-9}$ s~\cite{filippetti2014}. The pump pulse is then followed by a probe pulse that can be used to measure the AC conductivity of the new quasi-equilibrium system. To illustrate the potential usefulness of this technique for our purposes, consider a simple model where the recombination time $\tau_n$ is a constant. After a time $dt$, the number of carriers remaining in the conduction band will be $n_c(t+dt)=(1-\frac{dt}{\tau_n})n_c(t)$~\footnote{Note that in this very simple picture we have ignored the contribution of the holes to the AC transport properties, as well as any excitonic effects.}, so that
\be
n_c(t)=n_c(0)e^{-t/\tau_n}.
\ee
Thus, the delay time $t\sim\ln(n_c/n_c(0))$ provides an ideal control parameter for observing the quantized behaviour of the conductivity. We may compute the conductivity as a function of delay time using \eqref{eq:condfiniteT}, where the quasi-equilibrium chemical potential $\mu(t)$ is determined by the number equation
\be
n_c(T)=\frac{m}{2\pi}\int_0^{E_0}dE\sqrt{\frac{E_0}{E}}\frac{1}{e^{\beta(E-\mu)}+1}=n_c(0)e^{-t/\tau_n}.
\ee
The result is shown in Fig.~\ref{fig:delay}, where the first plateau, now as a function of delay time, is easily visible at sufficiently low temperatures. This plot is for a fixed impurity density. For cleaner systems, one would see more plateaus at a given temperature. This is again in contrast to the first Born approximation result which smoothly decays to zero. 
%fig. 6
%%%%%%%%%%%%%%%%%%%%%%%%%%%%%%%%%%%
%%%%%%%%%%%%%%%%%%%%%%%%%%%%%%%%%%%%%%%%%
\section{Self-Consistent Full Born approximation}\label{sec:SCFBA}
Until now, we have looked exclusively at the semiclassical transport features of low-density Rashba systems. Given the delicate nature of the dependence of these features on the density, one might be skeptical that they survive a fully quantum treatment. The purpose of this section is to address this question. We will focus exclusively on the zero-temperature limit throughout this section.
%%%%%%%%%%%%%%
\subsection{Single-particle properties}\label{subsec:single}
\begin{figure}[t]
	\centering
	\includegraphics[width=\columnwidth]{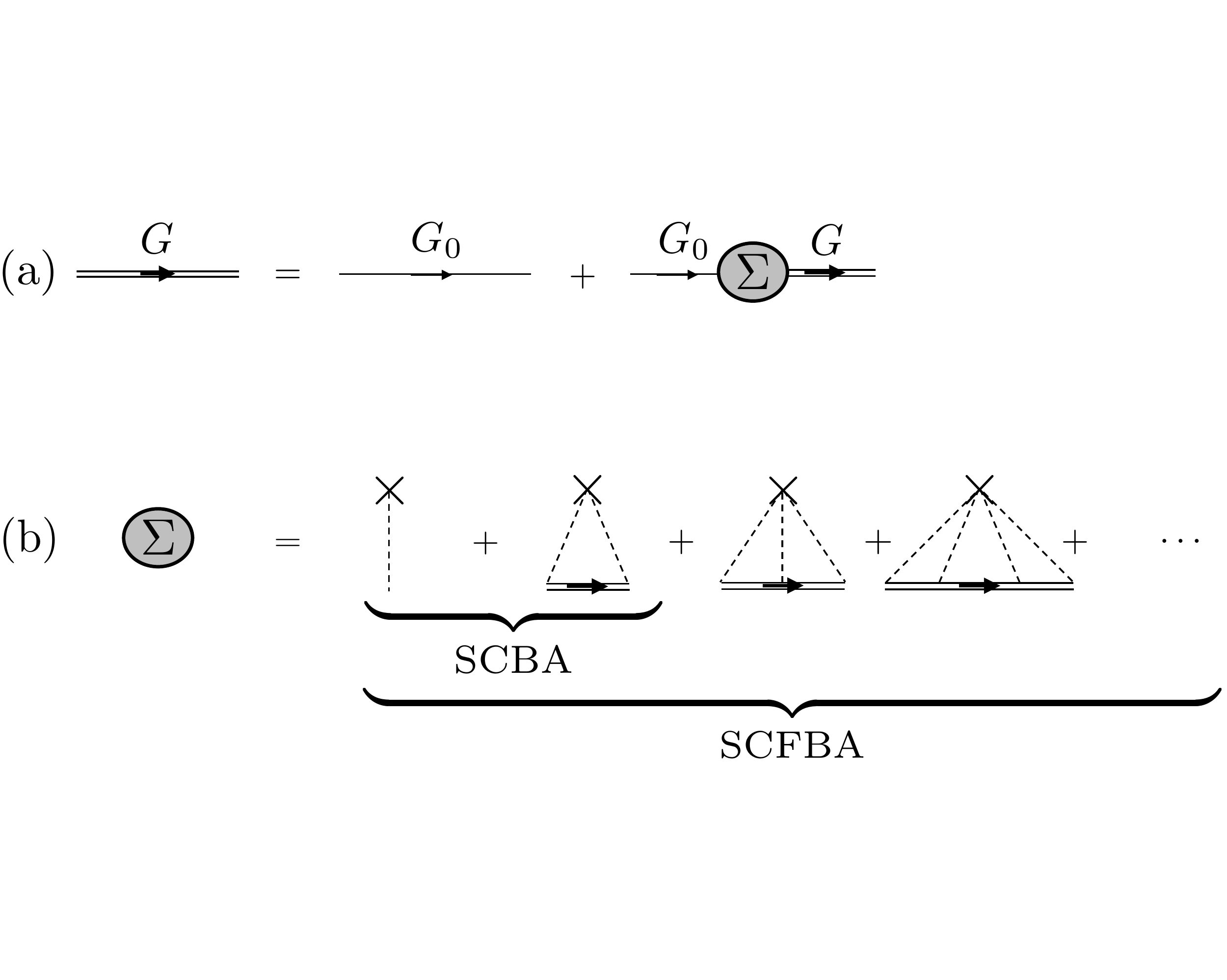}
\caption{(a) Dyson equation for the full Green's function $G$, where $G_0$ is the bare Green's function and $
\Sigma$ is the irreducible self-energy; (b) Diagrammatic expansion of the irreducible self-energy $\Sigma$ for impurity scattering, showing the truncation made in SCBA.  
}\label{fig:feynman1}
\end{figure}

We will utilize a self-consistent \emph{full} Born approximation (SCFBA). Note that this is different from the conventional self-consistent Born approximation (SCBA) used in~\cite{brosco2016, brosco2017}, in that the self-energy is determined self-consistently from the full $T$-matrix, and not simply the $T$-matrix in the first Born approximation. The distinction is illustrated in the diagrams in Fig.~\ref{fig:feynman1}. In the conventional SCBA, no diagrams in the self-energy have more than two impurity lines attached to a single vertex. In the SCFBA, one includes all non-crossed diagrams.
%fig. 7
We proceed with standard impurity averaging~\cite{mahan2013}. We start with $N_i$ uncorrelated impurities located at random positions $\b{R}_j$, described by the random potential
\be\label{eq:randomU}
U(\b{r})=\sum_{j=1}^{N_i}V(|\b{r}-\b{R}_j|).
\ee
The Green's function and self-energy are averaged over impurity positions. Since these positions only enter through phase factors $e^{i\b{q}\cdot\b{R}_j}$ in the Fourier transform of \eqref{eq:randomU}, the averaging induces a factor $n_i$ and a momentum-conserving delta function at each vertex. The series shown in Fig. \ref{fig:feynman1} (b) is precisely the Born series for the $T$-matrix with the addition of these self-averaging factors. Thus, we take the irreducible retarded self-energy in the helicity basis to be
\be
\Sigma_{\alpha\beta}(\b{k},E)=n_iT_{\alpha\beta}^{\b{k},\b{k}}(E).
\ee
The $T$-matrix satisfies the Born series,
\begin{eqnarray}\label{eq:Born}
&&T_{\alpha\beta}^{\b{k},\b{k}'}(E)=V_{\alpha\beta}(\b{k},\b{k}')\nonumber\\
&&+\sum_{\lambda\gamma}\int \frac{d^2\b{q}}{(2\pi)^2}\int \frac{d^2\b{q}'}{(2\pi)^2}V_{\alpha\lambda}(\b{k},\b{q})G_{\lambda\gamma}^{\b{q},\b{q}'}(E)T_{\gamma\beta}^{\b{q}',\b{k}'}(E),\nonumber\\
\end{eqnarray}
where $V_{\alpha\beta}(\b{k},\b{k}')$ is the matrix element of $V(\b{r})$ in the momentum-helicity basis. $G$ is the \emph{full} retarded Green's function in this basis, which obeys the Dyson equation, 
\begin{eqnarray}\label{eq:Dyson}
G_{\alpha\beta}(\b{k},E)&=&G^0_{\alpha\alpha}(\b{k},E)\delta_{\alpha\beta}\nonumber\\
&&+\sum_{\gamma}G^0_{\alpha\alpha}(\b{k},E)\Sigma_{\alpha\gamma}(\b{k},E)G_{\gamma\beta}(\b{k},E).\nonumber\\
\end{eqnarray}
Note that the impurity averaging procedure restores translation invariance, so that the Green's function is diagonal in momentum.
For circularly symmetric potentials, which have angular components
\be\label{eq:circularpotential}
V^l(k,k')=\int_0^{2\pi}\frac{d\theta}{2\pi}\int_0^\infty dr\;rV(r)J_0(|\b{k}-\b{k}'|r)e^{il\theta},
\ee
the low-energy $T$-matrix in the negative-helicity sector is independent of the magnitude of the momenta: 
\be
T_{--}^{\b{k}\b{k}'}=\sum_{l=-\infty}^\infty T^l(E)e^{il(\theta_{\b{k}}-\theta_{\b{k}'})}.
\ee 
This follows from the arguments made in Ref.~\cite{hutchinson2017}, which also hold for the SCFBA Green's function. This guarantees that the self-energy is independent of momentum in the same limit: $\Sigma_{--}(\b{k},E)\equiv\Sigma(E)$. By \eqref{eq:Dyson}, this also means that the Green's function is independent of $\theta_k$. The full Born series \eqref{eq:Born} is then solved by
\begin{eqnarray}
T^l(E)&=&\frac{1}{2}\bigg(\frac{V^l(k_0,k_0)(1-J_+^l+J_-^l)}{1-I_-^l-J_+^l+I_-^lJ_+^l-I_+^lJ_-^l}\bigg)\nonumber\\
&&+\frac{1}{2}\bigg(\frac{V^{l+1}(k_0,k_0)(1-J_+^l-J_-^l)}{1-I_-^l-J_+^l+I_-^lJ_+^l-I_+^lJ_-^l}\bigg),
\end{eqnarray}
where $I_\pm^l$ and $J_\pm^l$ are the integral contributions of the lower and upper helicity Green's functions, respectively,
\begin{eqnarray}
I^l_\pm=\int^\infty_0\frac{dq\;q}{4\pi}[V^l(k_0,q)\mp V^{l+1}(k_0,q)]G_{--}(q,E)\nonumber\\
J^l_\pm=\int^\infty_0\frac{dq\;q}{4\pi}[V^l(k_0,q)\pm V^{l+1}(k_0,q)]G_{++}(q,E).\nonumber\\
\label{eq:Il}
\end{eqnarray}

\begin{figure}[t]
	\centering
	\includegraphics[width=\columnwidth]{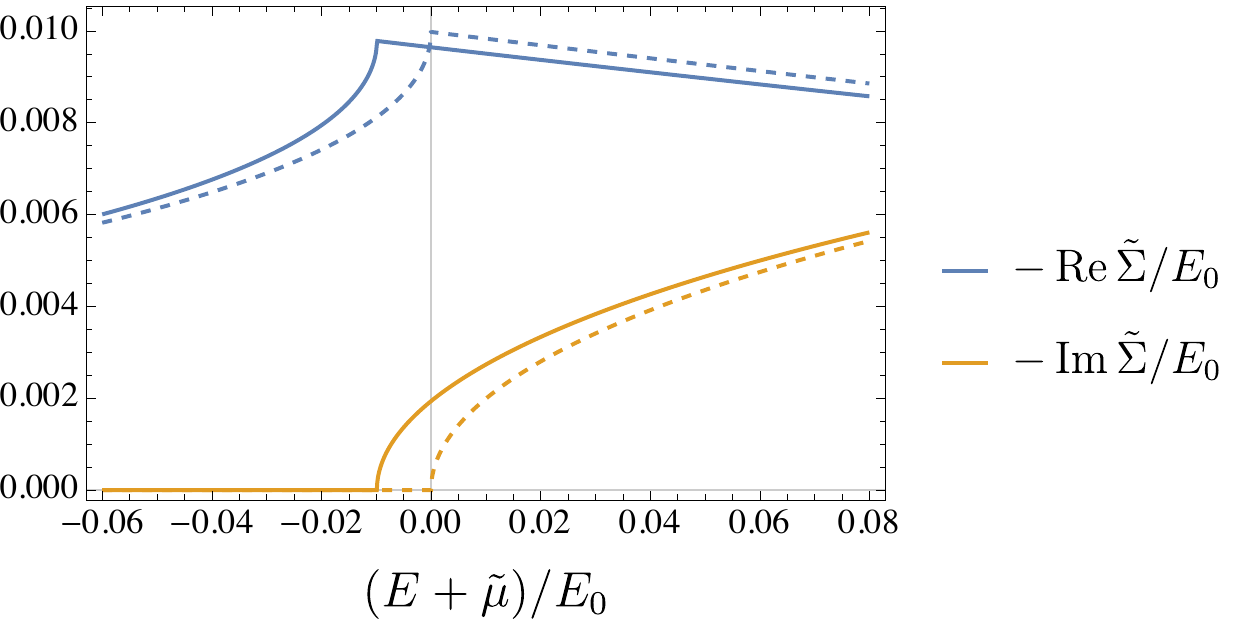}
\caption{Real and imaginary parts of the self-energy for the SCFBA (solid lines) and the full Born approximation (dashed lines). The two coincide in the clean limit. Here, the impurity density was chosen to be $n_i/n_0=0.016$. A cutoff of $\Lambda=0.5$ was used, although the self-energy in this regime is largely independent of this choice.
}\label{fig:selfplt}
\end{figure}

In the low-energy regime of interest to us, the integrand of $J_\pm^l$ is far from its poles in $q$ and we expect $J_\pm^l$ to be negligible. More precisely, let us impose a momentum cutoff $k_0\Lambda$ around the ring of degenerate states such that $\Lambda\ll1$, and then integrate from $k_0(1-\Lambda)$ to $k_0(1+\Lambda)$. 
%Over this range, the potential components $V^l(k_0,q)$ decay monotonically due to the Bessel function in \eqref{eq:circularpotential}, provided they are sufficiently short-ranged ($k_0R<\pi/2$). Thus we can construct the following bound for the integrand $f^l_\pm(q,E)$ of $J^l_\pm$:
%\be
%\bigg|f^l_\pm(q,E)\bigg|\leq\bigg|\frac{(1+\Lambda)[V^l(k_0,k_0(1-\Lambda))\pm V^{l+1}(k_0,k_0(1-\Lambda))]}{4\pi(E-\xi^+_q-\Sigma_{++}(E))}\bigg|
%\ee
In this range, $|J_\pm^l|\sim\Lambda\ll1$. 
The $T$-matrix is then determined entirely by the integral $I^l_-$:
\be\label{eq:lowETmat}
T^l(E)\approx\frac{\delta_l^*/m}{1-I^l_-}.
\ee
The integral $I_-^l$ depends on the Green's function component that satisfies the Dyson equation
\begin{eqnarray}\label{eq:greens}
G_{--}(k,E)&=&\bigg(G^0_{--}(k,E)^{-1}-\Sigma_{--}(E)\nonumber\\
&&-\frac{\Sigma_{-+}(E)G^0_{++}(k,E)\Sigma_{+-}(E)}{1-G^0_{++}(k,E)\Sigma_{++}(E)}\bigg)^{-1}.
\end{eqnarray}
The last term, containing the off-diagonal parts, is second order in the impurity density and will be ignored from now on. $I_-^l$ is derived in Appendix.~\ref{subsec:Ilm} to be 
\be
I_-^l\approx-i\frac{\delta_l^*}{z}-\frac{2\delta_l^*}{\pi\Lambda},
\ee
where
\be
z\equiv \sqrt{(E+\mu)/E_0-\Sigma(E)/E_0}.
\ee
We thus have the following self-consistency condition for the self-energy,
\be
\Sigma(E)=\frac{n_i}{m}\sum_{l=-\infty}^\infty\frac{\delta_l^*}{1+i\delta_l^*/z-\frac{2\delta_l^*}{\pi\Lambda}}.
\ee
Note that by expanding to lowest order in $n_i$, we get the self-energy corresponding to the full Born approximation as expected,
\be
\Sigma(E)\approx\frac{n_i}{m}\sum_{l=-\infty}^\infty\frac{\delta_l^*}{1+i\delta_l^*[(E+\mu)/E_0]^{-1/2}-\frac{2\delta_l^*}{\pi\Lambda}}.
\ee
It is conventional to absorb the lowest order self-energy term into the chemical potential. This amounts to redefining 
\begin{eqnarray}
\tilde{\Sigma}(E)&\equiv&\Sigma(E)-n_iV_{--}(k_0,k_0)\\
\tilde{\mu}&\equiv&\mu+n_iV_{--}(k_0,k_0).
\end{eqnarray}
The resulting self-energy is shown in Fig. \ref{fig:selfplt}.
%fig. 8

From \eqref{eq:greens} we may compute the spectral function $A(\b{k},E)=-2\im G_{--}(\b{k},E)$. It is a Lorentzian with an energy-dependent width given by
\be
1/\tau(E)=-2\im\Sigma(E)=n_i\sum_l\im T^l(E).
\ee
This is equivalent to the definition of the lifetime used in the Boltzmann description \eqref{eq:tau} due to the optical theorem for the low-energy $T$-matrix~\cite{hutchinson2017}:
\be\label{eq:tau2}
\im T^{\b{k}\b{k}'}_{--}(\theta=0)=-\frac{m}{2\pi}\int_0^{2\pi}d\theta\;|T^{\b{kk}'}_{--}|^2,
\ee
where $\theta$ is the angle between $\b{k}$ and $\b{k}'$.

\begin{figure}[t]
	\centering
	\includegraphics[width=\columnwidth]{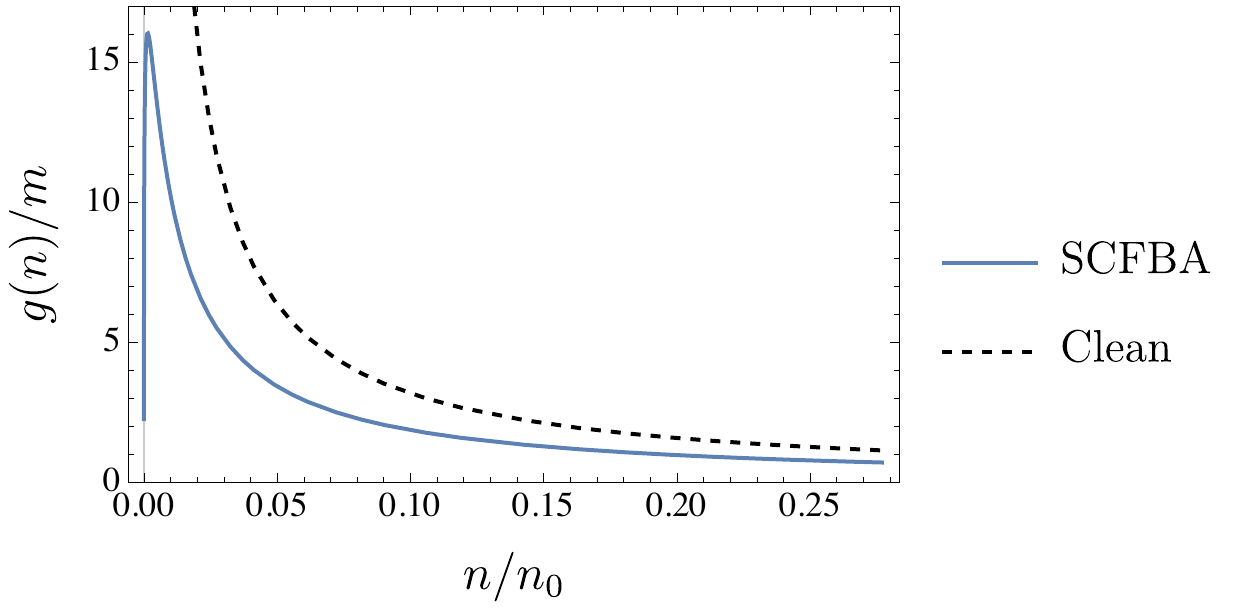}
\caption{Density of states as a function of electron density computed from the SCFBA for an impurity density of $n_i/n_0=0.016$, compared to the clean limit.}\label{fig:dos}
\end{figure}

From the spectral function, we obtain the density of states 
\begin{eqnarray}\label{eq:dos2}
g(E)&=&-\frac{\im\Sigma}{2\pi^2}\int_{k_0(1-\Lambda)}^{k_0(1+\Lambda)}\frac{dk\;k}{(E-\xi^-_k-\re\Sigma)^2+(\im\Sigma)^2},\nonumber\\
\end{eqnarray}
applying the same cutoff as before. This integral is similar to $I^l_-$ and is derived in Appendix.~\ref{subsec:dos}. The result is
\be
g(E)=\frac{m}{\pi}\re\bigg(\sqrt{\frac{E_0}{E+\tilde{\mu}-\tilde{\Sigma}(E)}}\bigg).
\ee
Note that in the clean limit, $\tilde{\Sigma}(E)\rightarrow0$, $\tilde{\mu}\rightarrow\mu$, and we recover the non-interacting density of states \eqref{eq:dos}. Integrating this up to the Fermi level gives the density, which we invert to obtain $g(n)$ as shown in Fig.~\ref{fig:dos}. As expected, disorder rounds the van Hove singularity in the density of states. 

A similar rounding of the density of states was found in the study of a 2D Rashba electron gas with delta-function impurities, for which an asymptotically exact solution is available in the low-energy limit~\cite{galstayn1998}. Here we are considering a more general situation making use of the universal behaviour of the $T$-matrix for arbitrary circularly symmetric, finite-range potentials \eqref{eq:lowETmat}.

%fig. 9
%%%%%%%%%%%%%%%%%%%%%
\subsection{Kubo conductivity}\label{subsec:Kubo}
\begin{figure}[t]
	\centering
	\includegraphics[width=\columnwidth]{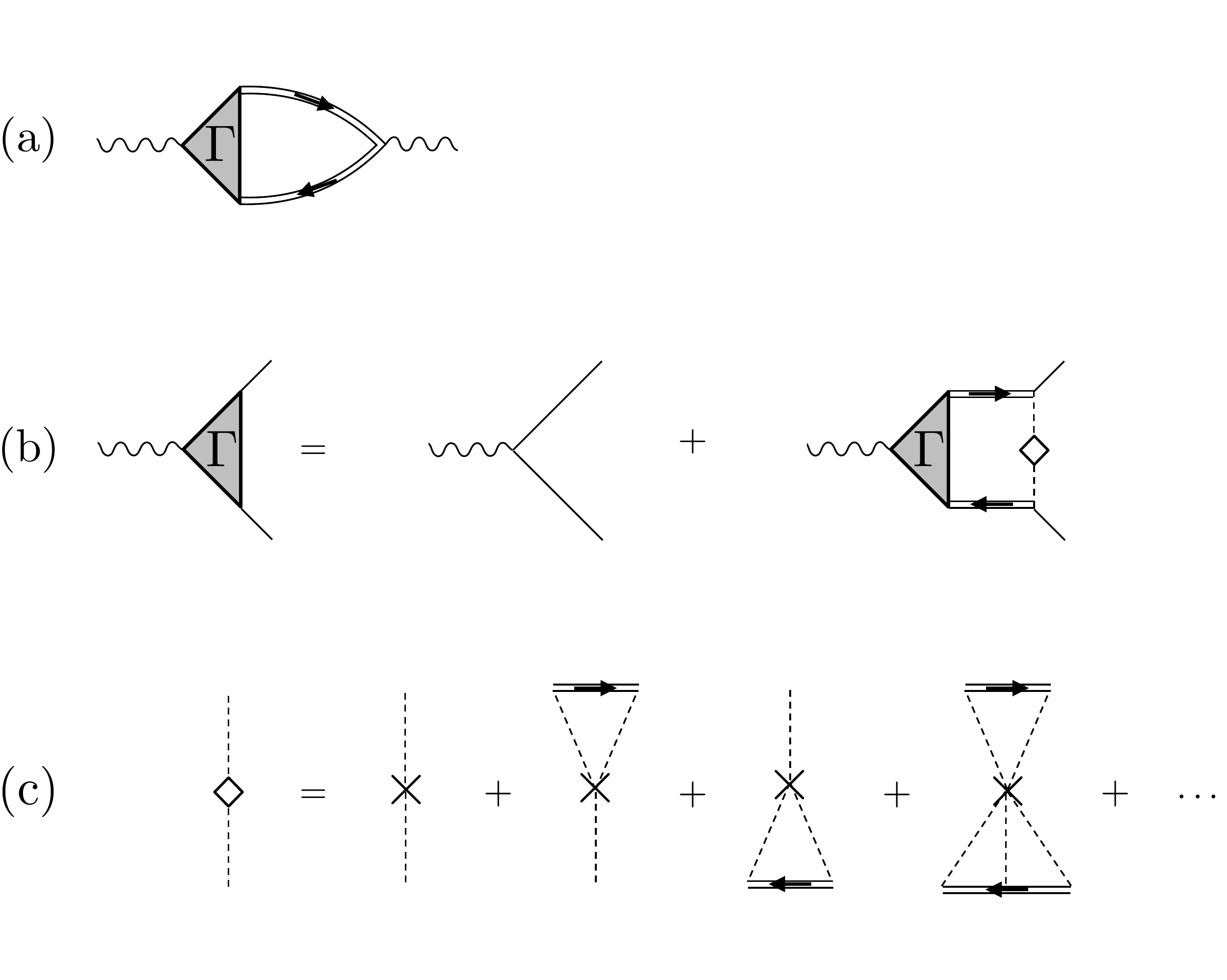}
\caption{(a) Conductivity bubble; (b) integral equation for the vertex part $\Gamma^{\rm{XR}}$. Here, the upper double line corresponds the Green's function $G^{\rm{X}}$, and the lower one corresponds to $G^{\rm{R}}$. (c) Born series for the $T$-matrix product. Double lines represent the SCFBA Green's function.}\label{fig:feynman2}
\end{figure}

\begin{figure}[t]
	\centering
	\includegraphics[width=\columnwidth]{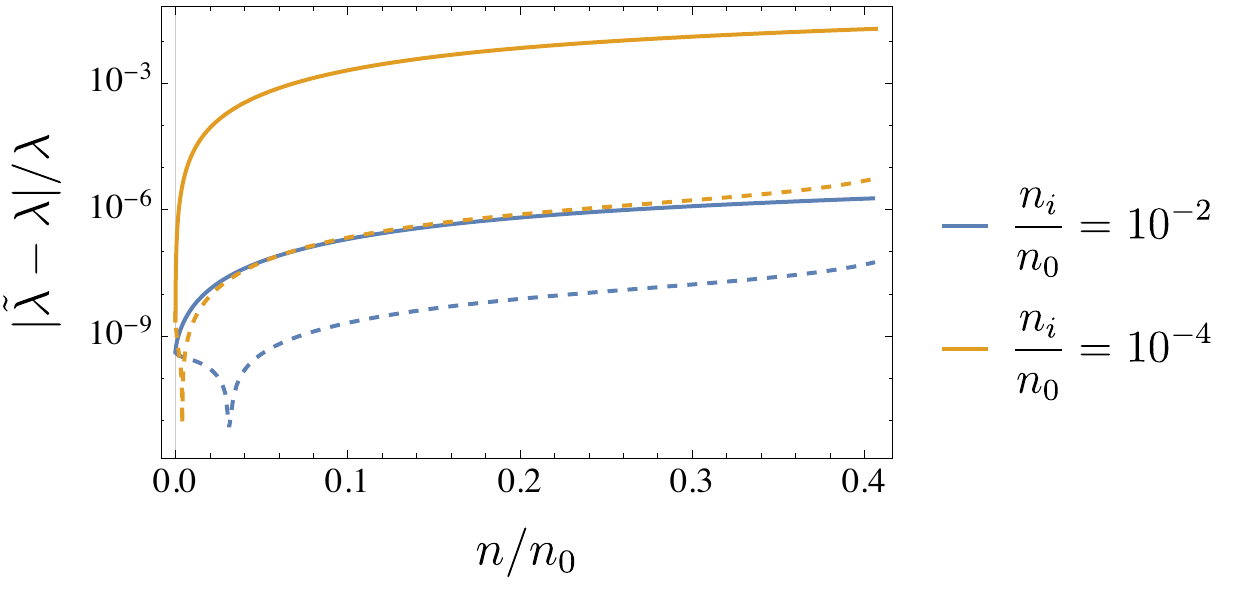}
\caption{Renormalized Rashba couplings $\tilde{\lambda}^{AR}$ (solid) and $\tilde{\lambda}^{RR}$ (dashed) relative to the bare coupling for two different impurity densities. The advanced-retarded coupling is slightly smaller than the bare coupling. The cusp seen in the retarded-retarded coupling occurs at $n\approx n_i$ and corresponds to a change in sign of $\tilde{\lambda}^{RR}-\lambda$. This coupling is slightly smaller than the bare value at electron densities below the impurity density, but becomes larger than the bare coupling above the impurity density.}\label{fig:relativelamplt}
\end{figure}

Using the Green's function and $T$-matrix derived in the previous section, we now look at the conductivity within linear response theory. In the Kubo formalism, the conductivity is given by
\be
\sigma^{\rm{DC}}=-e^2\lim_{\omega \rightarrow 0}\bigg(\frac{\im\Pi_{\rm{ret}}(\omega)}{\omega}\bigg),
\ee
where $\Pi_{\rm{ret}}(\omega)$ is the retarded current-current correlator shown diagrammatically in Fig.~\ref{fig:feynman2}(a).
This reduces to the standard expression
%, and given by the expression
%\begin{eqnarray}
%\Pi_{\rm{ret}}(\omega)&=&-\frac{1}{2\pi i}\int_{-\infty}^\infty d\epsilon f(\epsilon)[P(\epsilon+i\delta,\epsilon+\omega+i\delta)\nonumber\\
%&&-P(\epsilon-i\delta,\epsilon+\omega+i\delta)+P(\epsilon-\omega-i\delta,\epsilon+i\delta)\nonumber\\
%&&-P(\epsilon-\omega-i\delta,\epsilon-i\delta)].
%\end{eqnarray}
%One may check that $P(\epsilon-i\delta,\epsilon+i\delta)^*=P(\epsilon+i\delta,\epsilon-i\delta)=P(\epsilon-i\delta,\epsilon+i\delta)$, and that $\re(P(\epsilon+i\delta,\epsilon+\omega+i\delta))=\re(P(\epsilon-i\delta,\epsilon+\omega-i\delta))$. With these results, the conductivity takes the standard form
\be\label{eq:Kubocond}
\sigma^{\rm{DC}}=\frac{e^2}{2\pi}\int_{-\infty}^\infty dE\bigg(-\frac{\partial f}{\partial E}\bigg)\bigg[P^{\rm{AR}}(E)-\re P^{\rm{RR}}(E)\bigg],
\ee
where we have defined the advanced ($P^{\rm{AR}}(E)$) and retarded ($P^{\rm{RR}}(E)$) response functions via
\be
P^{\rm{XR}}(E)\equiv\int\frac{d^2\b{p}}{(2\pi)^2}\tr G^{\rm{X}}(p,E)\Gamma_0(\b{p})G^{\rm{R}}(p,E)\Gamma^{\rm{XR}}(\b{p},E),
\ee
with $X\in\{R,A\}$. Here, $G^{\rm{R}}$, and $G^{\rm{A}}$ are the retarded and advanced Green's functions. $\Gamma^{\rm{RR}}(\b{p},E)$, and $\Gamma^{\rm{AR}}(\b{p},E)$ are the retarded-retarded and advanced-retarded vertex parts, which satisfy the integral equation shown in Fig. \ref{fig:feynman2}(b). Lastly, $\Gamma_0(\b{p})\equiv\frac{\partial H}{\partial p_x}$ is the bare vertex. In the helicity basis, it is given by
\begin{eqnarray}
\Gamma_0(\b{p})=\frac{p_x}{m}\mathbb{I}-\lambda\cos\theta_\b{p}\sigma_z-\lambda\sin\theta_\b{p}\sigma_y.
\end{eqnarray}
Isotropy of the system allows us to just consider the $x$-component of the vertex part, corresponding to the longitudinal conductivity $\sigma^{\rm{DC}}_{xx}$. The integral equation for the vertex part in this basis is, in matrix notation,
\begin{eqnarray}
\Gamma^{\rm{XR}}(\b{p},E)&=&\Gamma_0(\b{p})\nonumber\\
&&+\int\frac{d^2\b{k}}{(2\pi)^2}T^{\b{p}\b{k}}(E)G^{\rm{X}}(k,E)\Gamma^{\rm{XR}}(\b{k},E)\nonumber\\
&&\times G^{\rm{R}}(k,E)T^{\b{k}\b{p}}(E).\label{eq:vertexpart}
\end{eqnarray}
%fig. 10
Since we are only interested in energies near the band bottom, we may neglect the contribution from the upper-helicity component of the Green's functions as they do not have any poles near those energies. We may then regard \eqref{eq:vertexpart} as a scalar equation in the lower-helicity sector. In this sector, $\Gamma_0(\b{p})=\frac{p_x}{m}-\lambda\cos\theta_{\b{p}}$, which motivates us to use the following ansatz for the renormalized vertex,
\be
\Gamma^{\rm{XR}}(\b{p},E)=\frac{p_x}{m}-\tilde{\lambda}^{\rm{XR}}(E)\cos\theta_\b{p}.
\ee
We have anticipated a renormalized Rashba coupling $\tilde{\lambda}^{\rm{XR}}(E)$ independent of momentum. The mass cannot be renormalized because the only term on the right-hand side of \eqref{eq:vertexpart} that depends on the magnitude of $\b{p}$ is the bare vertex. Expanding the $T$-matrix in circular harmonics again, Eq. \eqref{eq:vertexpart} reads
\begin{eqnarray}
\tilde{\lambda}^{\rm{XR}}&(E)&=\lambda-\frac{n_i}{2\pi}\sum_{ll'}T^l(E-i\delta)T^{l'}(E+i\delta)\nonumber\\
&\times&\int_0^{2\pi}\frac{d\phi}{2\pi}e^{i(l-l')\phi}(\cos\phi-\sin\phi\tan\theta_\b{p})\nonumber\\
&\times&\int_{k_0(1-\Lambda)}^{k_0(1+\Lambda)} dk\;k\bigg(\frac{k}{m}-\tilde{\lambda}^{\rm{XR}}(E)\bigg)G^{\rm{X}}(k,E)G^{\rm{R}}(k,E).\nonumber\\\label{eq:vertex2}
\end{eqnarray}
Our ansatz for the vertex part works because the mirror symmetry of the $T$-matrix [see Eq.\eqref{eq:Tmatsin}] guarantees that 
\be
\sum_{ll'}T^l(E-i\delta)T^{l'}(E+i\delta)\int^{2\pi}_0\frac{d\phi}{2\pi}e^{i(l-l')\phi}\sin\phi=0,
\ee
so that the $\theta_\b{p}$ dependence in \eqref{eq:vertex2} disappears. Using the lifetimes defined in \eqref{eq:tau} and \eqref{eq:tautr}, the renormalized coupling is
\be
\frac{\tilde{\lambda}^{\rm{XR}}(E)}{\lambda}=\frac{1+\frac{\delta}{4\pi k_0}(\frac{1}{\tau^{\rm{tr}}}-\frac{1}{\tau})P_2^{\rm XR}}{1+\frac{\delta}{4\pi m}(\frac{1}{\tau^{\rm{tr}}}-\frac{1}{\tau})P_1^{\rm XR}},
\ee
where
\begin{eqnarray}
P_1^{\rm XR}&\equiv&\int_{k_0(1-\Lambda)}^{k_0(1+\Lambda)} dp\; pG^{\rm{X}}(p,E)G^{\rm{R}}(p,E),\\
P_2^{\rm XR}&\equiv&\int_{k_0(1-\Lambda)}^{k_0(1+\Lambda)} dp\; \frac{p^2}{m}G^{\rm{X}}(p,E)G^{\rm{R}}(p,E).
\end{eqnarray}
The integrals for the advanced-retarded part may be computed analytically as shown in Appendix. \ref{subsec:AR} [Eqs. \eqref{eq:AR1}, \eqref{eq:AR2}]. The result is shown as a function of density in Fig. \ref{fig:relativelamplt}. At low density, the Rashba coupling renormalization is minimal.

%fig. 11

\begin{figure}[t]
	\centering
	\includegraphics[width=\columnwidth]{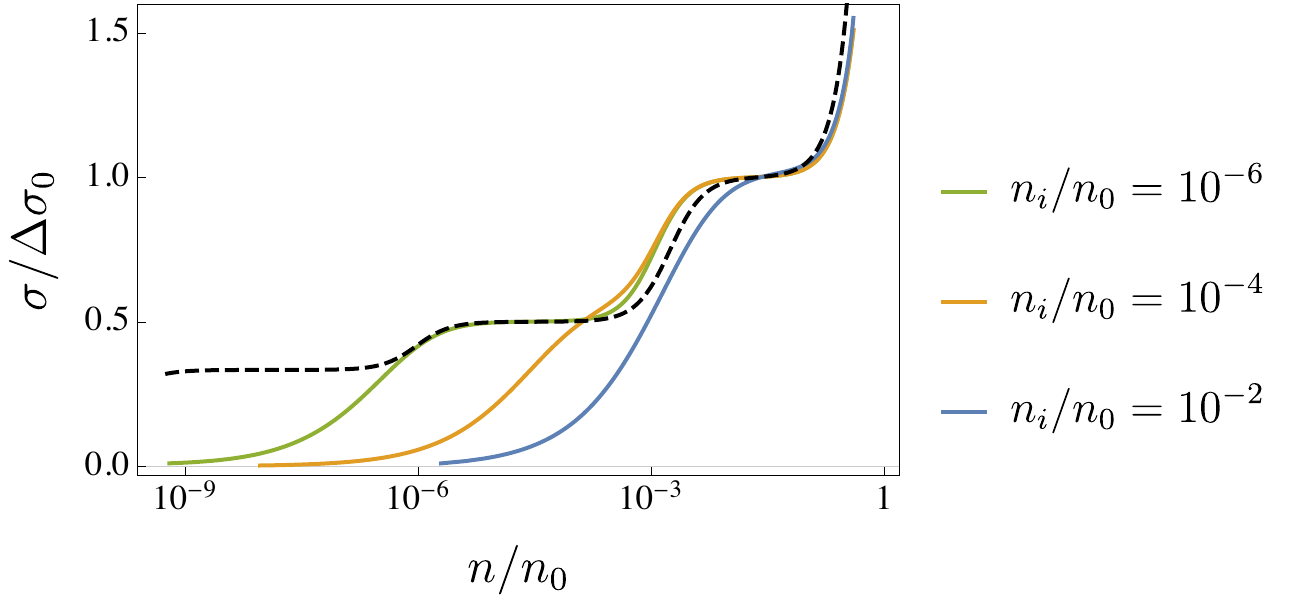}
\caption{Zero-temperature DC conductivity as a function of electron density computed numerically from the SCFBA for different impurity densities. We have normalized each curve by the impurity-density-dependent factor $\Delta\sigma_0\equiv\frac{e^2}{8\hbar}\frac{n_0}{n_i}$ [Eq.~\eqref{eq:DeltaSigma}]. For comparison, the Boltzmann result is shown with a dashed line. As shown in the text, the neglected crossing diagrams may become important for  $n<n_i$. Thus the range of validity of these curves is: $n/n_0>10^{-2}$ for the blue curve, $n/n_0>10^{-4}$ for the orange curve, and $n/n_0>10^{-6}$ for the green curve. In these ranges we see agreement with the Boltzmann result.}\label{fig:SCFBAcond}
\end{figure}

Having obtained the renormalized coupling, we may evaluate the response function,
\begin{eqnarray}
P^{\rm XR}(E)&=&\frac{1}{4\pi}\bigg(\int dp\; \frac{p^3}{m^2}G^{\rm X}(p,E)G^{\rm R}(p,E)\nonumber\\
&&-(\lambda+\tilde{\lambda}(E))\int dp\; \frac{p^2}{m}G^{\rm X}(p,E)G^{\rm R}(p,E)\nonumber\\
&&+\lambda\tilde{\lambda}(E)\int dp\; pG^{\rm X}(p,E)G^{\rm R}(p,E)\bigg).
\end{eqnarray}
Once again, the advanced-retarded integrals are evaluated analytically in Appendix.~\ref{subsec:AR}; we obtain
\begin{eqnarray}\label{eq:PAR}
P^{\rm AR}(E)&=&\frac{1}{\pi}(\tilde{\lambda}^{\rm AR}/\lambda-2)\nonumber\\
&&\times\bigg(\frac{2}{\Lambda}+\frac{\pi}{\im\tilde{\Sigma}}\re\sqrt{E_0[E+\tilde{\mu}-\tilde{\Sigma}(E)]}\bigg).\nonumber\\
\end{eqnarray}
It should be noted that the retarded-retarded part $P^{\rm RR}(E)$ only becomes important for electron densities below the impurity density. Above this density, the zero-temperature conductivity is well-approximated by
\be
\sigma^{\rm{DC}}\approx\frac{e^2}{2\pi}P^{\rm AR}(E).
\ee
Using \eqref{eq:PAR}, one can show that this reduces to the Boltzmann result. This is also clearly seen numerically in Fig. \ref{fig:SCFBAcond}, where the conductivity is computed from the full expression \eqref{eq:Kubocond} at zero temperature. We see that the prominent features of the DC conductivity found in the Boltzmann calculation (the drop near zero density, and the quantization on a log scale) survive in the fully quantum Kubo formula calculation as long as $n_i<n$. Note that this regime is consistent with the implicit assumption in the impurity-averaging process, namely that there is enough electron-electron interaction to cause decoherence between impurity-scattering events.

%fig. 12

\begin{figure}[t]
	\centering
	\includegraphics[width=\columnwidth]{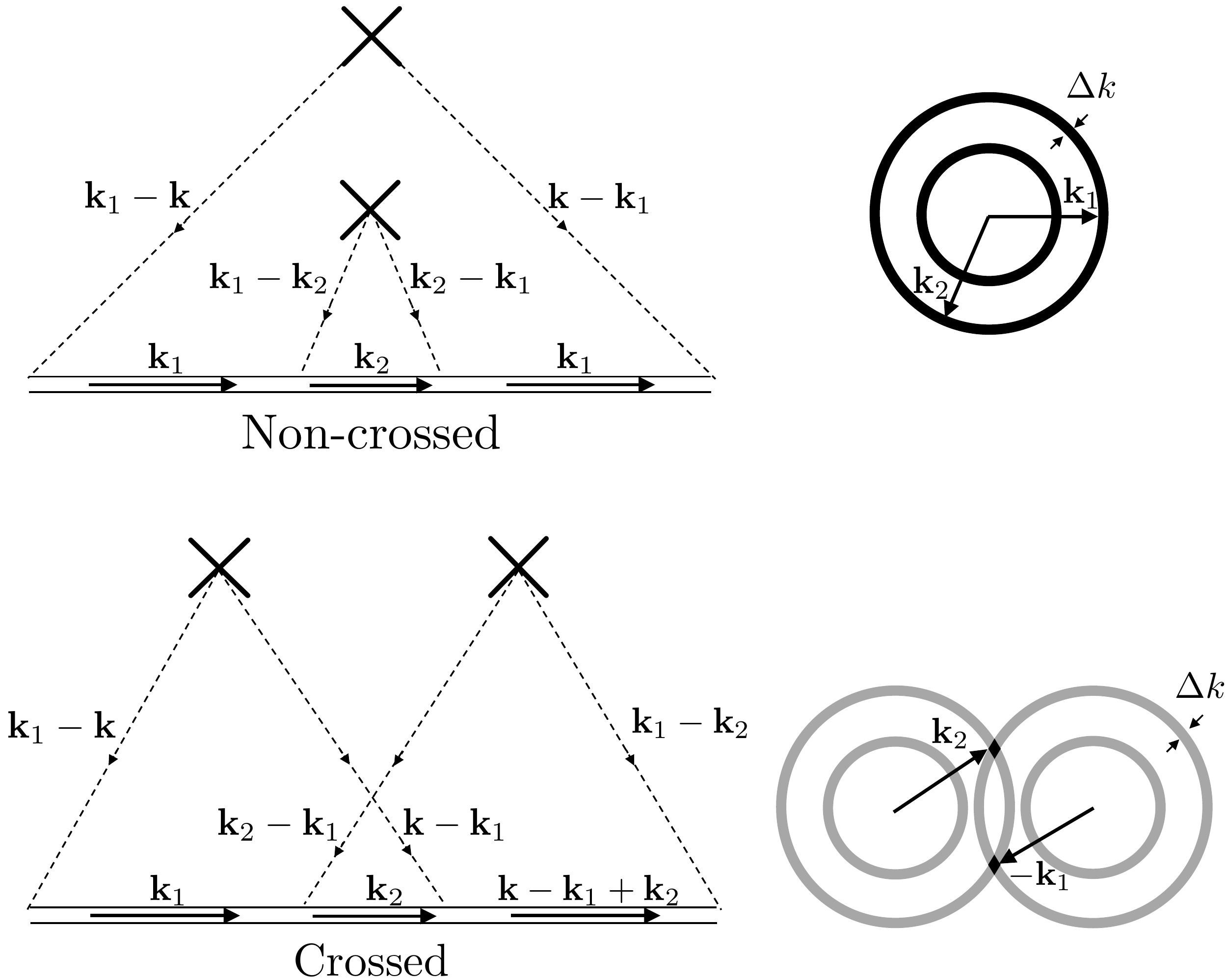}
\caption{Top: example of a non-crossed Feynnman diagram that contributes to the irreducible part of the SCFBA self-energy and the corresponding phase space for internal momenta. Both $\mathbf{k}_1$ and $\mathbf{k}_2$ can lie anywhere within the two black annuli defined by the Fermi surface. Bottom: a crossed diagram of the same order (second order in $n_i$, fourth order in the interaction), and one example of the corresponding phase space. Once $\b{k}_2$ is fixed, $-\b{k}_1$ is restricted to lie within the black diamonds.}\label{fig:feynman3}
\end{figure}

One might be skeptical about trusting the SCFBA in such a low density regime. First, Fermi liquid theory (resulting in Boltzmann transport) typically breaks down at ultra-low densities~\cite{bruus2004}. For another, the rapid drop seen in the conductivity as the density is lowered indicates a diverging scattering rate, and one might think that this could lead to interimpurity interference effects, such as weak antilocalization~\cite{golub2016}. As it turns out however, this is not the case, provided we focus on $n_i\ll n$, which is precisely the Boltzmann limit. To see this, recall that the only diagrams excluded from the SCFBA are the crossed diagrams, which give rise to quantum interference effects. An example of such a crossed diagram is shown in Fig. \ref{fig:feynman3}, compared to a non-crossed diagram of the same order. We know that the SCFBA spectral function is a Lorentzian with a width $1/\tau$ given in Eq. \eqref{eq:tau2}. At the Fermi level $E_F$, this corresponds to a smearing $(\Delta k)$ in momentum space that satisfies 
\be
\frac{[k_F\pm(\Delta k)]^2}{2m}-\lambda[k_F\pm(\Delta k)]\sim E_F+1/\tau,\label{eq:smear}
\ee
where $+$ and $-$ correspond to the $>$ and $<$ states respectively.  As shown in Ref.~\cite{galstayn1998}, the condition for crossed diagrams to be negligible is that $(\Delta k)\ll k_0$. We include this argument here for completeness. The result of Eq. \eqref{eq:smear} is that 
\be
(\Delta k)\sim k_0\delta\bigg(-1+\sqrt{1+\frac{2m}{(k_0\delta)^2}(1/\tau)}\bigg).
\ee
Returning to the diagrams in Fig. \ref{fig:feynman3}, we see that the internal momenta in the non-crossed diagram are independent, so that the phase space for this diagram is
\be
\Omega_{\rm{NC}}=[(2\pi k_<+2\pi k_>)(\Delta k)]^2=[4\pi k_0(\Delta k)]^2.
\ee
On the other hand, crossing diagrams have the restriction $|\b{k}_2+\b{k}-\b{k}_1|\in[k-(\Delta k), k+(\Delta k)]$, which means that once one momentum is fixed, the other is restricted to the intersection of four annuli. One possibility is shown in the bottom right of Fig.~\ref{fig:feynman3} with two intersections, but there are three other cases with four, six and eight intersections as well. Regardless, the phase space will be
\be
\Omega_{\rm{C}}\sim8\pi k_0(\Delta k)^3.
\ee
So we can neglect crossing diagrams (at least at this order), provided that
\be
\Omega_{\rm{C}}/\Omega_{\rm{NC}}\sim\frac{(\Delta k)}{2\pi k_0}\ll 1.
\ee
In the low-energy regime, this means that
\be\label{eq:NoWeak}
\frac{1}{E_0\tau}\ll4\pi^2.
\ee

%fig. 13

Let us now see if the SCFBA scattering rate meets this criteria. To be consistent with impurity averaging and the low-density approximations we have made, we should focus on
\be
\frac{n_i}{n_0}\ll\frac{n}{n_0}\ll1.
\ee 
In this case, we have seen that the scattering rate is well approximated by the low density Boltzmann result. From \eqref{eq:BoltzmannCondLowDensity}, we have
\be\label{eq:BoltzRate}
\frac{1}{E_0\tau}=\frac{e^2}{\pi}\bigg(\frac{n}{n_0}\bigg)\frac{1}{\sigma_{\rm DC}}.
\ee
If the conductivity drops to zero too rapidly as the electron density is lowered, one is not able to satisfy \eqref{eq:NoWeak}. But using Eq. \eqref{eq:plateaus}, as the index $l$ of the plateau increases, the conductivity decreases at a rate of
\be\label{eq:dsigmadl}
\frac{d\sigma_{\rm DC}}{d l}=-\frac{n_0}{n_i}\frac{e^2}{8l^2},
\ee
treating $l$ as a continuous variable and ignoring the detailed non-linear behaviour. Likewise, the density at each plateau transition is given by $n/n_0\approx\delta^*_l$, which for $\delta$-shell impurities \eqref{eq:dlstarDeltaShell} means
\be
\ln(n/n_0)=\ln\bigg(\frac{(k_0R)^2}{4}\bigg)l-2\ln(|l|!)+c,
\ee
and
\be
\frac{d(n/n_0)}{dl}=\frac{n}{n_0}\bigg[\ln\bigg(\frac{(k_0R)^2}{4}\bigg)-2\psi(|l|+1)\bigg],
\ee
where $c=\ln(mv_0R^2/2)$ and $\psi(x)$ is the digamma function. Now if $l$ is large, the conductivity will decay slowly according to \eqref{eq:dsigmadl}, and the $n/n_0$ pre-factor in \eqref{eq:BoltzRate} will ensure that the crossed diagrams are negligible. The only concern therefore is when $l$ is small. But in this case, $\psi(|l|+1)\approx-\gamma$, and $\ln(|l|!)\approx-\gamma l$, where $\gamma$ is the Euler-Mascheroni constant. The result is
\be
l^2=\bigg(\frac{\ln(n/n_0)-c}{\ln[\frac{(k_0R)^2}{4}]+2\gamma}\bigg)^2,
\ee
and
\be
\frac{d\sigma_{\rm DC}}{d(n/n_0)}=\frac{d\sigma_{\rm DC}}{dl}\frac{dl}{d(n/n_0)}\sim\frac{n_0/n_i}{n/n_0[\ln(n/n_0)-c]^2}.
\ee
Integrating with respect to $n/n_0$, we see that, roughly speaking, the conductivity changes with the density according to
\be
\sigma_{\rm{DC}}\sim\frac{n_0/n_i}{\ln(n/n_0)},
\ee
so that
\begin{eqnarray}
\frac{1}{E_0\tau}&\sim&(n/n_0)(n_i/n_0)\ln(n/n_0)\\
&&\ll (n/n_0)^2\ln(n/n_0)\\
&&\ll 1.
\end{eqnarray}
Thus we can trust the SCFBA result in the regime where the conductivity quantization is observed.
%%%%%%%%%%%%%%%%%%%%%%%%%%%%
\section{Conclusion}
The main result of this work was to show that the low-density conductivity due to impurity scattering in a 2D Rashba system takes a highly non-linear form that exhibits quantization as a function of the logarithm of the electron density. This unusual behaviour arises from the full non-perturbative low-energy $T$-matrix describing electron-impurity scattering near the ring minimum at the bottom of the Rashba conduction band. In the limit of a single impurity, this $T$-matrix was discussed in detail in Ref.~\cite{hutchinson2017}. 

It is clear that this highly degenerate band minimum is responsible for many unusual characteristics of low-energy transport. For one thing, the Fermi surface consists of two concentric circles with group velocities in opposite directions. At zero temperature, in the low-density limit, this produces an unconventional Drude-like expression for the conductivity that is controlled by the electron lifetime, as opposed to the usual transport time that appears in the conventional Drude formula. The transition from conventional to unconventional Drude transport as the density is lowered results in a non-linear conductivity as a function of density. This behaviour was first pointed out in Ref.~\cite{brosco2016}. The focus of our paper has been on the \emph{ultra}-low density regime, where the unconventional Drude conductivity becomes quantized. We showed this quantization within a semiclassical Boltzmann treatment (provided the full $T$-matrix is used in the scattering rate), as well as a fully quantum Kubo formula treatment, provided the electron density remains larger than the impurity density. 

The most important distinction between this and previous work is the use of the non-perturbative $T$-matrix in calculations. A $T$-matrix limited to the first Born approximation (or even a self-consistent first Born approximation) leads to qualitatively different transport phenomena. In particular, the first Born conductivity decays smoothly to zero with decreasing electron density, while the full $T$-matrix leads to a seemingly abrupt drop at zero density. For a Rashba semiconductor, we showed that this translates to a sharp drop in conductivity as the chemical potential passes through the band bottom, and that this drop retains significant weight at finite temperature.

We recognize that experimentally it is difficult to control the electron density with enough precision to access these ultra-low-density features. We have outlined a brief proposal of one way to overcome this difficulty in Rashba semiconductors. Namely, one could use a pump-probe approach in which the conductivity is measured as function of delay time between the two pulses. Such an approach allows one to use the logarithm of the carrier density as a control parameter. Of course, many technical issues would need to be addressed for such an experiment. One would need to carefully choose a 2D Rashba system with large splitting and Fermi level in the gap. The experiment would have to be done at very low ($\lesssim10$K) temperatures. Furthermore, the effect of hole carriers and excitons have not been addressed.

The last important point to emphasize is that our analysis is restricted to non-interacting systems. We recognize that at the low densities we are considering here, the effect of electron-electron interactions is enhanced, and one might expect Wigner crystallization to occur as a result. The electron-electron scattering process itself is dependent on a $T$-matrix that would likely contain unusual features similar to the impurity-scattering $T$-matrix considered here. Such a $T$-matrix may enhance or suppress the transport properties described in this paper or produce unique signals of its own. To say more would require explicit calculations that are beyond the scope of this paper. Instead, we can look for cases where we expect Fermi liquid theory to hold, outside the Wigner crystal regime. For one, it should be noted that many examples of Rashba 2D electron gases occur within gated samples. The presence of a metallic gate is expected to screen the Coulomb interaction, and the resulting short-range interaction may be insufficient to cause crystallization. Without knowing the details of the interaction it is hard to say more, though it should be noted that the unique low-energy density of states of the Rashba system may allow liquid crystal or anisotropic Wigner crystal phases to exist even for short-range interactions as described in Refs.~\cite{silvestrov2014, berg2012, ruhman2014}. Of course, the stability of these phases to disorder must be considered as well. Perhaps the simplest way to avoid the crystalline phase is to focus on temperatures above the melting point of the Wigner crystal. This occurs at a critical value of the dimensionless parameter $\Gamma=e^2\sqrt{\pi n}/(4\pi\epsilon_0k_BT)$, the ratio of potential and kinetic energies of a classical gas of electrons. In two dimensions, the melting point occurs around $\Gamma\approx130$~\cite{grimes1979, clark2009}. Using this number, we see that at the ultra-low densities considered in this paper, the Wigner crystal should melt at very low temperatures. For example, the density corresponding to the first plateau in Fig. \ref{fig:SCFBAcond} at $n/n_0\sim10^{-2}$ would be within the Fermi liquid phase for $T\gtrsim2$K. The second plateau at $n/n_0\sim10^{-5}$ corresponds to a melting temperature of $T\sim0.07$K. Now of course our SCFBA analysis was performed at zero temperature, but given the robustness of the non-perturbative transport effects to finite temperature in the Boltzmann treatment (Figs. \ref{fig:chempot}, \ref{fig:delay}), it is reasonable to assume that at least the first plateau would be observable at temperatures above the Wigner crystal melting point.

It is likely that these unusual transport features are not unique to Rashba systems. It would be interesting to determine exactly what aspects of this Hamiltonian are responsible for such non-linear behaviour. If the key aspect is the degenerate ring minimum in the band structure, then such features could also be observed in materials with pure Dresselhaus spin-orbit coupling~\cite{dresselhaus1955}. If the key aspect is the topology of the Fermi sea, then these features could appear in higher dimensional systems as well. Indeed, if the same quantization occurs in three-dimensional systems, then the group of candidate materials for experimental observation would be enlarged significantly.

%%%%%%%%%%%%%%%%%%%%%%%%%%%
\acknowledgements

We would like to thank David Purschke for his useful insights into pump-probe measurements. J.H. was supported by NSERC and Alberta Innovates - Technology Futures (AITF). J.M. was supported by NSERC grant \#RGPIN-2014-4608, the Canada Research Chair Program (CRC), the Canadian Institute for Advanced Research (CIFAR), and the University of Alberta.

%%%%%%%%%%%%%%%%%%%%%%%%%%%%%%%%%%%%%%%%
%%%%%%%%%%%%%%%%%%%%%%%%%%%%%%%%%%%%%%
%%%%%%%%%%%%%%%%%%%%%%%%%%%%%%%%%%%%
%%%%%%%%%%%%%%%%%%%%%%%%%%%%
%%%%%%%%%%%%%%%%%%%%%%
\appendix

\section{Symmetry of $T^l(E)$}\label{app:symm}
Besides rotation symmetry, which allows us to expand the $T$-matrix in circular harmonics with coefficients $T^l(E)$, the Rashba $T$-matrix for circular impurity potentials  is also symmetric under reflections in the $y-z$ plane. In the spin basis, this means
\be
\sigma_xT(M_x(\b{k},\b{k}'))\sigma_x=T(\b{k},\b{k}'),
\ee
where $M_x$ maps $k_x$ to $-k_x$. Transforming to the helicity basis, this condition becomes
\be
(-ie^{-i\phi_{\b{k}}}\sigma_z)T(M_x(\b{k},\b{k}'))(ie^{i\phi_{\b{k}'}}\sigma_z)=T(\b{k},\b{k}').
\ee
Both sides may be expanded in circular harmonics,
\be
\sum_le^{-i(l+1)(\phi_{\b{k}}-\phi_{\b{k}'})}\sigma_zT^l(E)\sigma_z=\sum_le^{il(\phi_{\b{k}}-\phi_{\b{k}'})}T^l(E).
\ee
Shifting $l\rightarrow-l-1$, we get
\be
\sigma_zT^{-l-1}(E)\sigma_z=T^{-l}(E).
\ee
For the lower helicity component $T_{--}$, this means
\be
T^{l-1}(E)=T^{-l}(E).
\ee
Note that this condition guarantees detailed balance in the Boltzmann scattering rate \eqref{eq:fermirule}, since
\begin{eqnarray}
|T^{\b{k}'\b{k}}|^2&=&\bigg|\sum_lT^l(E)e^{il(\phi_{\b{k}'}-\phi_{\b{k}})}\bigg|^2\\
&=&\bigg|\sum_lT^{-l}(E)e^{il(\phi_{\b{k}}-\phi_{\b{k}'})}\bigg|^2\\
&=&\bigg|\sum_lT^{l-1}(E)e^{il(\phi_{\b{k}}-\phi_{\b{k}'})}\bigg|^2\\
&=&\bigg|\sum_lT^{l}(E)e^{il(\phi_{\b{k}}-\phi_{\b{k}'})}\bigg|^2|e^{i(\phi_{\b{k}}-\phi_{\b{k}'})}|^2\\
&=&|T^{\b{k}\b{k}'}|^2.
\end{eqnarray}

Another important consequence of this symmetry is the identity
\be
\int_0^{2\pi}\frac{d\phi}{2\pi}\sin\phi|T^{\b{k}\b{k}'}|^2=0,\label{eq:Tmatsin}
\ee
where $\phi$ is the angle between $\b{k}$ and $\b{k}'$. This follows from expanding the left-hand side in circular harmonics to get
\begin{eqnarray}
&&\frac{1}{2i}\sum_{l}\bigg(T^l(E)^*T^{l-1}(E)-T^l(E)^*T^{l+1}(E)\bigg)\nonumber\\
&=&\frac{1}{2i}\sum_{l}\bigg(T^{-l}(E)^*T^{l}(E)-T^{-l}(E)^*T^{l}(E)\bigg)\\
&=&0,
\end{eqnarray}
where we used mirror symmetry and shifted the summation index in each term.
\section{Derivation of important integrals}\label{app:Il}

%(for the delta-shell, $|\sum_{l=-\infty}^\infty\delta_l^*|<mV_0R(e^{(k_0R)^2/4}-1/2)$).

%If $\Sigma\rightarrow0$ as $E+\mu\rightarrow0$, we can always choose $|E+\mu|<<\Lambda^2$ so that $|z|<\Lambda$. If $\Sigma\rightarrow\infty$ as $E+\mu\rightarrow0$,
%\be
%G_{++}(q,E)=\frac{1}{E-4E_0+\mu-n_iT_{++}(q,E)}+\mathcal{O}(\Lambda).
%\ee
%Since $E-\mu<E_0$, we can construct the following bound on $J^l_\pm$:
%\be
%|J_\pm^l|\leq\int_{k_0(1-\Lambda)}^{k_0(1+\Lambda)}dq\bigg|\frac{k_0[V^l(k_0,k_0)\pm V^{l+1}(k_0,k_0)]}{4\pi(3E_0+n_iT^{q,q}_{++}(E))}+\mathcal{O}(\Lambda)\bigg|.
%\ee
%If the $T$-matrix in the denominator is singular at $E\rightarrow0$, we can make $J_\pm$ as small as we like in the low energy limit, if not, we will choose $n_i$ to be small enough that $T$-matrix term is negligible. In either case, we can establish the bound
%\be
%|J_\pm^l|\leq\Lambda\frac{k_0[V^l(k_0,k_0)\pm V^{l+1}(k_0,k_0)]}{6\pi E_0}+\mathcal{O}(\Lambda^2) <<1.
%\ee 
\subsection{$I^l_-$}\label{subsec:Ilm}
We first derive the low-energy form of $I_-^l$ [equation \eqref{eq:Il}]. This is the integral that governs the energy dependence of the $T$-matrix. We use the cutoff scheme $k_0(1-\Lambda)<k<k_0(1+\Lambda)$ with $\Lambda\ll1$, and letting $\epsilon=(q-k_0)/k_0$, we have
\be
I^l_-=\frac{m}{2\pi}\int^\Lambda_{-\Lambda}d\epsilon f(\epsilon;E),
\ee
where
\be
f(\epsilon;E)\equiv\frac{(1+\epsilon)[V^l(k_0,k_0(1+\epsilon))+V^{l+1}(k_0,k_0(1+\epsilon))]}{(E+\mu)/E_0-\epsilon^2-\Sigma(E)/E_0}.
\ee
The integrand $f(\epsilon;E)$ has two poles located at $\epsilon=\pm z$, where
\be
z\equiv\sqrt{(E+\mu)/E_0-\Sigma(E)/E_0}.
\ee
%\begin{eqnarray}
%\epsilon_\pm&=&\frac{\pm1}{\sqrt{2}}[(|z|+\re z)^{1/2}+i(|z|-\re z)^{1/2}],
%\end{eqnarray}
%where we have defined $
We can perform this integral by considering a semicircular contour of radius $\Lambda$ through the upper half-plane, so that
\be
I^l_-=mi\;\underset{\epsilon=z}{\Res}f(\epsilon;E)-\frac{mi\Lambda}{2\pi}\int_0^\pi d\phi\; e^{i\phi} f(\Lambda e^{i\phi};E).
\ee
Keeping terms in the numerator of $f(\Lambda e^{i\phi};E)$ at lowest order in $\Lambda$, we have
\be\label{eq:Iint}
I^l_-=mi\;\underset{\epsilon=z}{\Res}f(\epsilon;E)-\frac{i\Lambda\delta^*_l}{\pi}\int^\pi_0\frac{d\phi\; e^{i\phi}}{z^2-\Lambda^2e^{2i\phi}},
\ee
where we have discarded the $\mathcal{O}(\Lambda)$ terms in the expansion of $V^l(k_0,k_0(1+\Lambda e^{i\phi}))$ since these are suppressed by an additional factor of $k_0R$ that we take to be small. 
Each pole is located a distance $z$ from the origin. Whether or not our contour encloses the $\epsilon=+z$ pole depends on the low-energy behaviour of the (unknown) self-energy. We now show that the contour must enclose this pole. We will make use of the following identity:
%\be\label{eq:intcases}
%\int^\pi_0\frac{d\phi e^{i\phi}(1+\Lambda e^{i\phi})}{z^2-\Lambda^2e^{2i\phi}}=\begin{cases}
%\frac{2i}{z\Lambda}\Arctanh{(\frac{\Lambda}{z})} & \Lambda<|z|\\
%-\frac{2i}{z\Lambda}\Arctanh{(\frac{z}{\Lambda}})-\frac{\pi}{\Lambda} & \Lambda>|z|\\
%\end{cases},
\be\label{eq:intcases}
\int^\pi_0\frac{d\phi\; e^{i\phi}}{z^2-\Lambda^2e^{2i\phi}}=\begin{cases}
\frac{2i}{z\Lambda}\Arctanh{(\frac{\Lambda}{z})}, & \Lambda<|z|,\\
-\frac{2i}{z\Lambda}\Arctanh{(\frac{z}{\Lambda}}), & \Lambda>|z|,\\
\end{cases}
\ee
$\Arctanh(x)$ being the principal value of $\arctanh(x)$. 
Suppose for contradiction that the $\epsilon=+z$ pole lies outside our contour. Then, we have the first of the two cases in \eqref{eq:intcases}, and the residue is zero:
\be
I_-^l=\frac{2\delta_l^*}{\pi z}+\mathcal{O}(\Lambda^3/z^4).
\ee
This would mean $|I_-^l|<\Lambda\ll1$, so that  $T^l(E)\approx\delta_l^*/m$, using \eqref{eq:lowETmat}. In that case the self-energy would be
\be
\Sigma(E)=\frac{n_i}{m}(mV_{--}(k_0,k_0)).
\ee
%The sum over $\delta_l^*$ is a number of the order of the impurity strength $\lesssim 1$. 
The prefactor must satisfy $n_i/m\ll \Lambda^2E_0$ so that the average impurity spacing is much larger than the inverse of the momentum cutoff scale. This implies that $|\Sigma(E)/E_0|\ll \Lambda^2$ (provided the impurity strength is not too large). But in the low-energy regime we are considering, $(E+\mu)/E_0\ll \Lambda^2$ as well, and so $|z|\ll \Lambda$, which is a contradiction. Thus the $\epsilon=+z$ pole contributes to \eqref{eq:Iint}, and using the second case of \eqref{eq:intcases}, we have
\begin{eqnarray}
I_-^l&\approx&-\frac{i\delta^*_l}{z}+\frac{i\Lambda\delta^*_l}{\pi}\bigg(\frac{2i}{z\Lambda}\Arctanh(z/\Lambda)\bigg)\nonumber\\
&=&-i\frac{\delta_l^*}{z}-\frac{2\delta_l^*}{\pi\Lambda}+\mathcal{O}(|z|^2/\Lambda^3).
\end{eqnarray}

%%%%%%%%%%%%%%%%
\subsection{Density of states}\label{subsec:dos}
We may apply the same contour to evaluate the integral in the density of states. From \eqref{eq:dos2},
\begin{eqnarray}
g(E)&=&-\frac{m}{\pi^2}\im\tilde{\Sigma}/E_0\nonumber\\
&&\times\int^\Lambda_{-\Lambda}\frac{d\epsilon(\epsilon+1)}{((E+\tilde{\mu}-\re\tilde{\Sigma})/E_0-\epsilon^2)^2+(\im\tilde{\Sigma}/E_0)^2}.\nonumber\\
\end{eqnarray}
The $\epsilon$ term in the numerator is odd, so we need only evaluate the integral
\be
g(E)=-\frac{m}{\pi^2}\frac{\im\tilde{\Sigma}}{E_0}\int^\Lambda_{-\Lambda}\frac{d\epsilon}{(a-\epsilon^2)^2+b^2},
\ee
where $a\equiv(E+\tilde{\mu}-\re\tilde{\Sigma})/E_0$ and $b\equiv\im\tilde{\Sigma}/E_0$. The integrand now has two poles in the upper half-plane, $\epsilon=\mp z_\pm\equiv\mp\sqrt{a\pm ib}$, both with magnitude $(a^2+b^2)^{1/4}$. By the same reasoning as before, these poles must be contained within the semicircle of radius $\Lambda$ and therefore contribute residues to the integral. The integral over the semicircle is given by
\begin{eqnarray}
&&i\Lambda\int^\pi_0\frac{d\phi\; e^{i\phi}}{(a-\Lambda^2e^{2i\phi})^2+b^2}\nonumber\\
&=&\frac{-\Lambda}{2b}\bigg(\int^\pi_0\frac{d\phi\; e^{i\phi}}{z_+^2-\Lambda^2e^{2i\phi}}-\int^\pi_0\frac{d\phi\; e^{i\phi}}{z_-^2-\Lambda^2e^{2i\phi}}\bigg)\\
&=&\frac{i}{b}\bigg[\frac{1}{z_+}\Arctanh(z_+/\Lambda)-\frac{1}{z_-}\Arctanh(z_-/\Lambda)\bigg],\nonumber\\
\end{eqnarray}
using \eqref{eq:intcases} again. The result is zero to order $|z|^2/\Lambda^3$. Thus we are just left with the residue contribution,
\begin{eqnarray}
g(E)&=&\frac{m}{\pi^2}\frac{\im\tilde{\Sigma}}{E_0}\frac{\pi}{2b}\bigg(\frac{1}{\sqrt{a+ib}}+\frac{1}{\sqrt{a-ib}}\bigg)\\
&=&\frac{m}{\pi}\re\bigg(\sqrt{\frac{E_0}{E+\tilde{\mu}-\tilde{\Sigma}(E)}}\bigg).
\end{eqnarray}
%%%%%%%%%%%%%
\subsection{Advanced-retarded integrals}\label{subsec:AR}
It turns out that all the integrals that enter the advanced-retarded part of the conductivity are simply higher moments of the density of states integral and can be solved analogously. We will look at the first three moments, denoted $P_1$, $P_2$, $P_3$. From \eqref{eq:dos2}, we immediately see that
\begin{eqnarray}
P_1&\equiv&\int_{k_0(1-\Lambda)}^{k_0(1+\Lambda)} dp\;pG^A(p,E)G^R(p,E)\\
&=&\frac{-2\pi^2}{E_0\im\tilde{\Sigma}}g(E).\label{eq:AR1}
\end{eqnarray}
Likewise, 
\begin{eqnarray}
P_2&\equiv&\int_{k_0(1-\Lambda)}^{k_0(1+\Lambda)} dp\;\frac{p^2}{m}G^A(p,E)G^R(p,E)\\
&=&\frac{k_0^3}{mE_0^2}\int^\Lambda_{-\Lambda}\frac{d\epsilon\epsilon^2}{(a-\epsilon^2)^2+b^2}+\lambda P_1.\nonumber\\
\end{eqnarray}
This time the integration over the semicircle gives
\begin{eqnarray}
&&i\Lambda^3\int^\pi_0\frac{d\phi\; e^{3i\phi}}{(a-\Lambda^2e^{2i\phi})^2+b^2}\nonumber\\
&=&\frac{i}{b}[z_-\Arctanh(z_-/\Lambda)-z_+\Arctanh(z_+/\Lambda)]\nonumber\\
&\approx&2/\Lambda.
\end{eqnarray}
Adding the residue contribution gives
\begin{eqnarray}
P_2&=&\lambda P_1-\frac{4}{\lambda}\bigg(\frac{2}{\Lambda}+\frac{\pi}{\im\tilde{\Sigma}}\re(\sqrt{(E+\tilde{\mu})/E_0-\tilde{\Sigma}(E)})\bigg).\nonumber\\\label{eq:AR2}
\end{eqnarray}
Lastly, the third moment can be obtained from the first two:
\begin{eqnarray}
P_3&\equiv&\int_{k_0(1-\Lambda)}^{k_0(1+\Lambda)} dp\;\frac{p^3}{m^2}G^A(p,E)G^R(p,E)\\
&=&4\int^\Lambda_{-\Lambda}\frac{d\epsilon(\epsilon+1)^3}{(1-\epsilon^2)^2+b^2}\\
&=&\lambda(3P_2-2\lambda P_1).
\end{eqnarray} 
%%%%%%%%%%%%%%%%%%%
\bibliography{RashbaConductivity}
\end{document}